# Calibration of MAJIS (Moons And Jupiter Imaging Spectrometer): I. On-ground setup description and characterisation

Mathieu Vincendon,[1] Pierre Guiot,[1] Benoît Lecomte,[1] Mathieu Condamin,[1] François Poulet,[1] Antoine Arondel,[1] Julien Barbay,[1] John Carter,[1] Simone De Angelis,[2] Cydalise Dumesnil,[1] Gianrico Filacchione,[2] Paolo Haffoud,[1] Jérémie Hansotte,[1] Yves Langevin,[1] Pierre-Louis Mayeur,[1] Giuseppe Piccioni,[2] Cédric Pilorget,[1,3] Eric Quirico,[4] and Sébastien Rodriguez[5]

[1])*Institut d'Astrophysique Spatiale, Université Paris-Saclay, CNRS, Bâtiment 121, 91400 Orsay, France*
[2])*INAF-IAPS, Istituto di Astrofisica e Planetologia Spaziali, via del Fosso del Cavaliere, 100, 00133, Rome, Italy*
[3])*Institut Universitaire de France, Paris, France*
[4])*Institut de Planétologie et d'Astrophysique (IPAG), UMR 5274, CNRS, Université Grenoble Alpes, Grenoble, France*
[5])*Institut de Physique du Globe de Paris, CNRS-Université Paris-Cité, Paris 75005, France*

(*Electronic mail: mathieu.vincendon@universite-paris-saclay.fr)

(Dated: 21 February 2025)

The visible and infrared Moon And Jupiter Imaging Spectrometer (MAJIS), aboard the JUpiter ICy Moons Explorer (JUICE) spacecraft, will characterize the composition of the surfaces and atmospheres of the Jupiter system. Prior to the launch, a campaign was carried out to obtain the measurements needed to calibrate the instrument. The aim was not only to produce data for the calculation of the radiometric, spectral, and spatial transfer functions, but also to evaluate MAJIS performance, such as signal-to-noise ratio and amount of straylight, under near-flight conditions. Here, we first describe the setup implemented to obtain these measurements, based on five optical channels. We notably emphasize the concepts used to mitigate thermal infrared emissions generated at ambient temperatures, since the MAJIS spectral range extends up to 5.6 µm. Then, we characterize the performance of the setup by detailing the validation measurements obtained before the campaign. In particular, the radiometric, geometric, and spectral properties of the setup needed for the inversion of collected data and the calculation of the instrument's calibration functions are presented and discussed. Finally, we provide an overview of conducted measurements with MAJIS, and we discuss unforeseen events encountered during the on-ground calibration campaign.

## NOMENCLATURE

BB   Black Body
DN   Digital Number
FOV   Field Of View
GEO   GEOmetric (or spatial) calibration items
IAS   Institut d'Astrophysique Spatiale
IFOV   Instantaneous Field Of View
IR   InfraRed channel
IS   Integrating Sphere
ITF   Instrument Transfert Function
JUICE   JUpiter ICy moons Explorer
MAJIS   Moon And Jupiter Imaging Spectrometer
M*   Plane Mirror No.*
OA*   Off-Axis conic mirror No.*
OB   Optical Bench (of the calibration setup)
OH   Optical Head (of MAJIS)
OP   Optical Path (of the calibration setup)
PM*   Parabolic Mirror No.*
QTH   Quartz Tungsten Halogen lamp
RAD   RADiometric calibration items
SIMBIO-SYS   Spectrometer and Imaging for MPO Bepi-Colombo Integrated Observatory SYStem
SPE   SPEctral calibration items
TVC   Thermal Vacuum Chamber
VISNIR   VISible and Near-InfraRed channel

## I. INTRODUCTION

### A. The MAJIS instrument

The *Moon And Jupiter Imaging Spectrometer* (MAJIS[1]) is an instrument on-board the space mission *Jupiter Icy Moons Explorer* (JUICE) of the European Space Agency. JUICE was launched on April 14, 2023, and is scheduled to arrive around the Jupiter system in 2031. One of the primary objectives of MAJIS is to map the composition of ices, minerals, and organic matter at the surface of Ganymede, Callisto, and Europa. JUICE will notably be the first probe to orbit a natural satellite other than the Moon (Ganymede). MAJIS will also observe the atmosphere and rings of Jupiter and its other satellites.

MAJIS spatial image construction is primarily based on the principle of push broom imagers. One spatial dimension is constructed by successive acquisitions during the along-track motion of JUICE on its trajectory. The other, across-track dimension is obtained at once with a slit perpendicular to the probe trajectory. The slit is associated with a detector line; detectors are thus 2D, as the other dimension enables spectral acquisition. In addition, MAJIS possesses a scan mirror that makes it possible to modify the boresight by $\pm 4°$ in the along-track (across-slit) direction. It will be used to scan distant objects or to compensate for incompatibilities between the along-track speed and the required acquisition time during close flybys.

The Field of View (hereafter FOV) of MAJIS is 59 mrad (3.4°) along-slit and 150 µrad across-slit (Table I). Along-slit, the FOV is divided into 400 Instantaneous Fields of View (hereafter IFOV) of 150 µrad. Trade-offs during the selection of the optical design and detectors of MAJIS lead to use detectors with a pixel size two times smaller compared to the needed size. Hence, MAJIS uses the concept of "nominal pixel"[1], obtained by binning two physical pixels on the detector. The spatial sampling thus corresponds to one IFOV, i.e. one nominal pixel, i.e., two physical pixels; it is equal to 150 µrad, which translates to a spatial sampling of 75 m at the surface of Ganymede during the orbital phase at 500 km altitude. Spatial oversampling at sub-IFOV with unbinned pixels may also be used during the mission.

TABLE I. Summary of MAJIS main characteristics needed to design the calibration setup[1].

| Parameter | Value |
| --- | --- |
| Along-slit Field of View (FOV) | ± 1.7° |
| Spatial sampling (IFOV) | 150 µrad |
| Number of nominal (binned) spatial pixels | 400 |
| Spectral range VISNIR | [0.49 - 2.35] µm |
| Spectral range IR | [2.27 - 5.56] µm |
| VISNIR spectral sampling | $\Delta\lambda \approx 3.7$ nm |
| IR spectral sampling | $\Delta\lambda \approx 6.5$ nm |
| Number of nominal (binned) spectral pixels | 508 |
| Pixel size (unbinned) | 18 µm |
| Nominal pixel size (binned) | 36 µm |
| Focal length | 240 mm |
| Entrance pupil dimensions | 84 mm × 57 mm |
| Entrance pupil equivalent diameter | 75 mm |
| System equivalent f-number | 3.2 |
| Scan mirror dimensions | 121 mm × 71 mm |
| Along-track scan capability | ± 4° |

MAJIS spectral range covers part of the visible and infrared ranges. Acquisitions are based on two channels: VISNIR (0.49 to 2.35 µm) and IR (2.27 to 5.56 µm) (Table I). Both channels first share a common optical path through a three-mirrors anastigmatic telescope with a 75 mm aperture equivalent diameter (the pupil is a rounded rectangle with dimensions 84 mm x 57 mm) and a 3.2 f-number (focal length divided by aperture equivalent diameter), until slit entrance[2]. The signal is then separated by a dichroic filter toward two grating spectrometers for wavelengths dispersion and two detectors for acquisition. The average spectral sampling is about $\Delta\lambda = 3.7$ nm for VISNIR and $\Delta\lambda = 6.5$ nm for IR, with 508 nominal pixels covering the full spectral range (the spectral sampling varies slightly along-slit and with wavelength[1,3]). Similarly to the spatial dimension, physical pixels on the detector are also binned by default in the spectral dimension to obtain nominal spectral pixels (sometimes called "spectels"). Acquisition with spectral oversampling (use of unbinned spectral pixels) can also be implemented with MAJIS[1].

### B. Calibration objectives

The MAJIS on-ground calibration campaign aimed at gathering data before launch to determine the radiometric, geometric and spectral response of the instrument in conditions compliant with those of the relevant planetary environment. This is a standard procedure for visible and infrared imaging or point spectrometers operating from planets' orbits or surfaces[4–14]. We list in Table II the MAJIS characteristics that were to be evaluated during the ground calibration campaign. Objectives are divided in three main categories: "RAD", "GEO" and "SPE" for the radiometric, geometric (or spatial) and spectral items respectively. Table II also describes the accuracy with which MAJIS parameters had to be determined. These accuracy requirements have notably been defined in such a way as to be able to verify the agreement between MAJIS's actual performance and the desired performance based on the scientific objectives[1]. They were also based on standards used in previous calibration campaigns of similar instruments[4,7].

For radiometry, the meaning of this accuracy is ambiguous and must be clarified. The Instrument Transfer Function (hereafter ITF), which provides the link between the output signal in digital numbers (hereafter DN) and the incoming radiance from objects observed by MAJIS, had to be determined with an absolute accuracy better than 20% and a relative accuracy better than 1%. This 1% relative accuracy is applied both in the spectral and spatial dimensions. Spatially, the 1% refers to the standard deviation of the distribution of MAJIS nominal pixels output values for a uniform source. Spectrally, the 1% relates to spectral depths over 10 $\Delta\lambda$ wide spectral ranges. For the straylight, the $10^{-5}$ contrast corresponds to the ratio of signal that should be detectable out-FOV (up to 10° away) over the near-saturation signal in-FOV. These characteristics had to be evaluated for various MAJIS operative configurations (thermal environment, integration time, etc.). These configurations are specified in Table III.

### C. Heritage

MAJIS ground calibration measurements were mainly acquired in the calibration facility of the *Institut d'Astrophysique Spatiale* (hereafter IAS) in Orsay, France. Additional measurements useful for MAJIS calibration were also obtained prior to the IAS campaign at the *Leonardo* compagny[1,3,15] but will not be discussed here.

The IAS facility main hall is an ISO8 cleanroom regulated to a temperature of 20°C and a humidity of 50%, which holds a large thermal vacuum chamber (hereafter TVC) located under an ISO5 laminar flow box. The facility also includes an optical bench previously used for the calibration of other imaging spectrometers such as OMEGA/Mars Express and VIRTIS/Rosetta in the 1990s / early 2000s[4,16], and then SIMBIO-SYS/Bepicolombo in 2015[7]. SIMBIO-SYS is a suite of three channels, two visible cameras and a visible-infrared hyperspectral imager "VIHI". In the SIMBIO-SYS configuration, the instrument was mounted on a mobile hexapod inside the TVC, while light sources were fixed and located outside the TVC in the optical bench[7]. This SIMBIO-SYS/Bepicolombo configuration was the starting point for the development of a new calibration setup adapted to MAJIS/JUICE.

SIMBIO-SYS/VIHI and MAJIS are indeed not alike as their respective targets, Mercury and Jupiter, receive a solar flux that differs by a factor of nearly $2 \times 10^2$, with cascading consequences. Mercury's high surface temperatures lead to limit the SIMBIO-SYS/VIHI wavelength range to $\lambda \leq 2$ µm where reflected sunlight is not drowned in surface thermal emission, while MAJIS will observe up to 5.6 µm. Consequently, the MAJIS detector must be cooled to about 90K, while 220 K was sufficient for SIMBIO-SYS/VIHI[17], and MAJIS optical head (hereafter OH) must be cooled to about 130 K (Table III). In addition, the ambient 20°C temperature of the facility does not create a significant parasitic thermal background at 2 µm, while this is no longer true beyond 4 µm. This latter point is worsened by the fact that the lower solar flux at Jupiter requires the MAJIS detector to be more sensitive to photons in general and, thus, to thermal background photons as well.

The objectives of the IAS calibration campaign were also different for both instruments: while the aim for MAJIS was to conduct a full calibration of the whole instrument, it was limited to a cross-calibration between channels already calibrated for SIMBIO-SYS[7,9,10,17–19].

TABLE II. Overview of MAJIS initial ground calibration objectives. (I)FOV refers to the (Instantaneous) Field of View of MAJIS and $\Delta\lambda$ to the spectral sampling (see section I A: $\Delta\lambda/5$ corresponds to 0.7 nm and 1.3 nm for VISNIR and IR respectively, and IFOV is 150 µrad). The "1% relative" accuracy and the "$10^{-5}$ contrast" are defined in the text of section I B.

| Item | Definition | Description | Accuracy |
|---|---|---|---|
| RAD-1 | Instrument transfer function (ITF) | Ratio between MAJIS output signal (digital numbers "DN") and incoming source radiance ($Wm^{-2}\mu m^{-1}sr^{-1}$) | Absolute 20% |
|  | Flat field | ITF variability over the FOV | 1% relative |
|  | Linearity and saturation | Linearity deviation of the ITF as a function of flux, and saturation levels | 1% relative |
| RAD-2 | Straylight | Quantify in-FOV signal caused by out-FOV illumination | $10^{-5}$ contrast up to $10°$ |
| GEO-1 | Boresight | Central viewing direction of pixels | $\leq$ IFOV/2 (75 µrad) |
|  | Field of view (FOV) | Angular size of the IFOV (single pixel) and FOV (whole slit) | $\leq$ IFOV/5 (30 µrad) |
| GEO-2 | Spatial point spread functions | Pixel and slit response functions to a point source | $\leq$ IFOV/5 (30 µrad) |
| SPE-1 | Spectral calibration | Absolute central wavelength of all pixels | $\leq \Delta\lambda/5$ |
|  | Spectral smile | Wavelength deviation along-slit | $\leq \Delta\lambda/5$ |
| SPE-2 | Spectral response function | Spectel response function to a monochromatic source | $\leq \Delta\lambda/5$ |
| SPE-3 | Samples | Identification of spectral bands on solid and gaseous samples | |

TABLE III. MAJIS operative parameters explored during the on-ground calibration. Calibration items are defined in Table II.

| Parameter | Value | Main relevant calibration items |
|---|---|---|
| IR detector temperature | 88 K to 100 K | RAD-1, SPE-23 |
| Optical head (OH) temperature | 126 K to 137 K | RAD-12, GEO-12, SPE-123 |
| Scan mirror across FOV pointing | $-4°$ to $+4°$ | RAD-12, GEO-12 |
| Integration time | 85 ms to 4 s | RAD-1 |

Overall, the requirements for MAJIS were thus more demanding in terms of extent of the wavelength range, cooling needs, and number of characteristics to be evaluated, compared to the inherited setup finally used for SIMBIO-SYS. We detail in the next two sections these requirements.

### D. Thermal requirements

We show in Figure 1 pre-calibration simulations of MAJIS signal while looking at ambient temperature optics or surfaces of the optical bench. The model used to elaborate these simulations is detailed in[20]: it incorporates optics from the optical bench and TVC, and MAJIS theoretical optical transmission and quantum efficiency. We consider the lowest (thus most favorable) integration time foreseen during calibration (see Table III). These simulations show that the contribution of transparent/reflective optics at ambient temperature was predicted to reach 50% of detector full well capacity at wavelengths 4.5 to 5 µm. They also illustrate that the direct observation of a black body surface at ambient temperature was foreseen to saturate MAJIS signal for all wavelengths greater than 3.5 to 4 µm (Figure 1).

As a consequence, thermal background mitigation was required for the development of the optical paths dedicated to the IR channel characterization, in particular for the assessment of radiometry (item RAD-1 of Table II). This came in addition to the requirements regarding the thermal environment of MAJIS in the TVC, which were much more demanding compared to SIMBIO-SYS as discussed in section I C (MAJIS IR detector and optics should be cooled at temperatures $\approx 90$ K and $\approx 130$ K respectively, see Table III). Overall, the development of a thermal environment adapted to MAJIS was one of the main challenges for the update of the calibration setup. Implemented solutions are presented in section II B.

### E. Spectral requirements

Other major modifications were related to the significantly extended IR wavelength range of MAJIS compared to SIMBIO-SYS/VIHI (up to 5.6 µm vs 2 µm). One of the key items of MAJIS on-ground calibration was the absolute radiometry (RAD-1 in Table II) which had to be determined over the whole FOV and spectral range. In order to make data analysis easier, it was decided to add to the setup two large, full-FOV flat field light sources able to cover both VISNIR and IR channels. In order to facilitate the handling of potential systematic biases during data analysis, it was decided that each of these two sources should cover entirely one channel (VIS-NIR or IR, respectively) with an overlap of about 0.5 µm over the other channel (i.e., $\approx 10\%$ of the whole spectral range).

Potential spectral perturbations associated with absorptions caused by water and carbon dioxide present in the ambient air led to specific requirements for these radiometric sources. The former SIMBIO-SYS optical bench was flushed with nitrogen. The detection threshold of humidity probes is typically 1%. We show in Figure 2 a simulation of the spectral transmission through a gas of dominant neutral nitrogen composition, mixed with a remaining contamination of ambient air containing water vapor (50% humidity) and carbon dioxide. The mix is such that the residual humidity is 1%. The optical path is $\approx 10$ m long, a value representative of the longest paths of the former SIMBIO-SYS bench[7]. In these conditions, water vapor and carbon dioxide produce several absorption bands greater than 1%, up to 5%, over the MAJIS spectral range (Figure 2). Thus, any uncontrolled variations of the amount of air in the bench below the 1% humidity threshold could have led to spectral fluctuations non compliant with the "1% relative" specification of RAD-1 (see Table II and text of section I B). As a consequence, it was decided to limit the optical path for radiometry channels to less than 2 m in the flushed bench. As discussed in the next section (section II A), one of the two radiometry channels was ultimately placed entirely under vacuum in the TVC, while the other was arranged at the closest possible optical distance of the TVC entrance window.

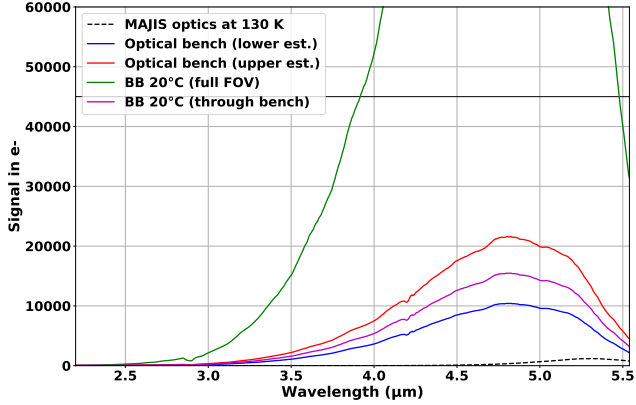

FIG. 1. Pre-calibration simulations of expected MAJIS signal resulting from the parasitic thermal flux of the optical bench under different assumptions. All optical surfaces are at ambient (20°C) temperature. The lowest foreseen nominal integration time (85 ms) is considered. Explored optical bench configurations are based on the inherited SIMBIO-SYS setup[7] (see section I C), which notably relies on the use of a main collimator with a 10 f-number[4]. Simulations include transparent or reflective surfaces with $\varepsilon = 0.03$ (mirrors, windows, etc.) and/or black body surfaces with $\varepsilon = 1$ (bench walls, source shutters, etc.) Green: black body directly in the field of view of MAJIS. Purple: black body placed at the focal plane of the collimator of the bench. Red: upper estimate of optics contribution with 5 optics directly observed by MAJIS. Blue: lower estimate of optics contribution with 2 optics directly observed by MAJIS and 3 optics placed at collimator focal plane. Dotted line: Thermal contribution of MAJIS internal optics for a 130 K optical head (OH). MAJIS saturation level is approximated with the horizontal black line at 45000 electrons (e-) per physical (unbinned) pixel. Details of the model used to performed these simulations are provided in[20].

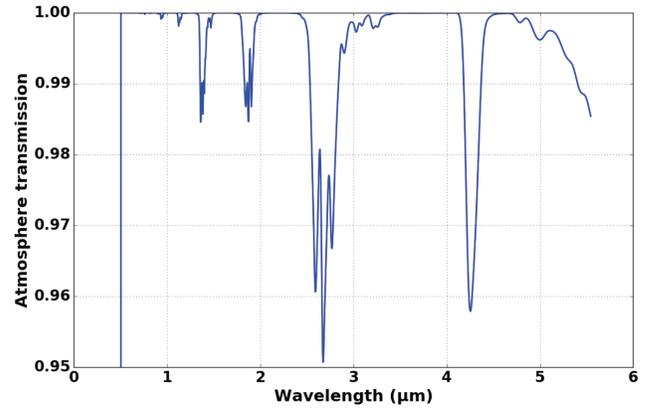

FIG. 2. Simulation of potential residual water vapor and carbon dioxide spectral features in a nitrogen flushed optical bench. The spectrum is generated using HITRAN[21]. A mix of gas at a total pressure of 1 *bar*, composed of 98% nitrogen and 2% air with a 50% humidity, is considered. This mix provides a total humidity of 1%, corresponding to the probe detection threshold. Data are convoluted to the expected spectral resolution of MAJIS and shown over the full MAJIS spectral range (Table I). An optical path of 10.5 m through the gas is simulated, as this length corresponds to the longest paths used previously in the bench[7]. Several water and/or carbon dioxide features exceed 1% depth, with the deepest absorptions near 2.7 µm reaching 5%.

## II. SETUP DESCRIPTION

### A. Global architecture

A first schematic representation of the setup is provided in Figure 3. The setup was divided into two parts: a 7 m long TVC (diameter: 2 m) to simulate the space environment with pressure down to $10^{-7}$ mbar[4], and an external Optical Bench (hereafter OB) for most light sources and optical components. MAJIS was located in the TVC. The optical connection between the TVC and the optical bench consisted of a calcium-fluoride window previously characterized[22].

The OB was composed of three cavities (Figure 4): a main rectangular structure with opaque walls inherited from the SIMBIO-SYS/BepiColombo setup[7], a new cubic unit with opaque walls added between the TVC window and the main rectangular structure, and a new transparent box adjacent to a side of the rectangular structure. All cavities of the OB were flushed with nitrogen to remove atmospheric gas. Four probes were placed in the OB to monitor humidity, with a lower detection limit of 1%. The new cubic unit was added in order to accommodate a VISNIR radiometric source at short distance from the TVC window to comply with requirements of section I E. The new transparent box was added to hold the monochromator that was previously in the ambient air.

The test beams used to calibration MAJIS were obtained through the use of several light sources and optical elements. Five main "Optical Paths" (hereafter OP) have been defined and numbered. OP can share common light sources or optics. These OP are briefly introduced below and subsequently described in details in the next sections:

- OP1 produced a spatially limited spot with wavelength selection capabilities. It was used for the geometric and spectral calibration.

- OP2 was designed to create a powerful polychromatic spot of a few tens of IFOV used for the radiometry, in particular straylight characterization.

- OP3 provided a radiometric flat field for the VISNIR channel. It was located in the new cubic unit of the OB.

- OP4 was equivalent to OP3 but for the IR channel. It was the only OP located inside the TVC.

- OP5 used a light source of OP2 to illuminate a samples carrier and direct reflected light toward MAJIS.

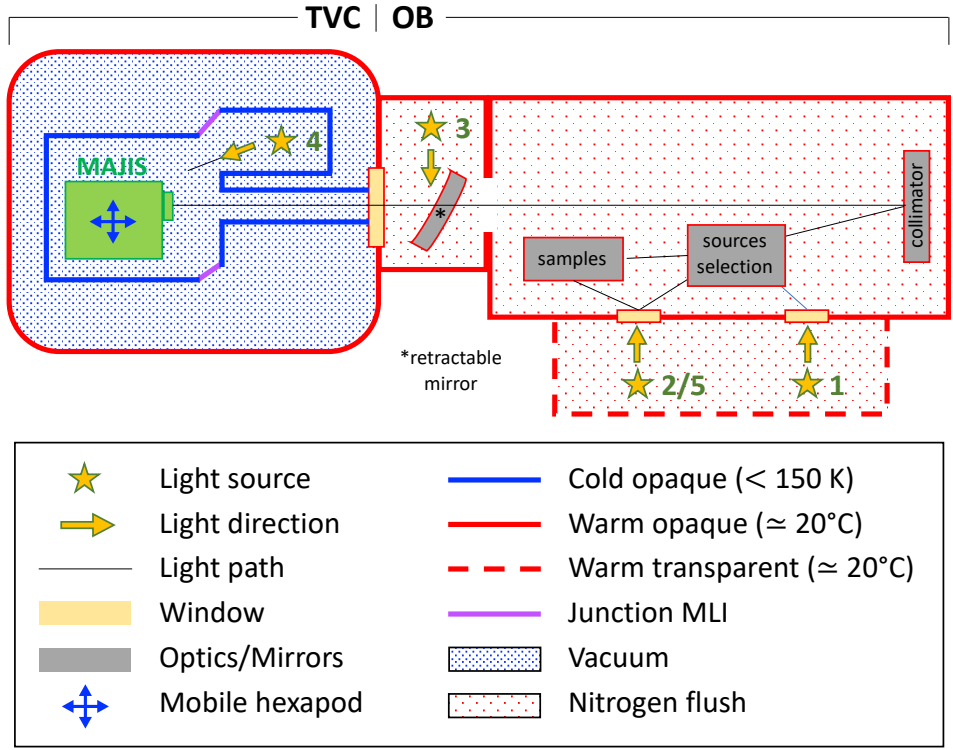

FIG. 3. Schematic overview of the setup architecture with the thermal vacuum chamber (TVC) on the left and the external optical bench (OB) on the right. Cold/warm and opaque/transparent surfaces are indicated. Numbers from 1 to 5 associated with light sources refer to "optical path" (OP) numbers detailed in section II A.

TABLE IV. Optical Paths (OP) used to fulfil calibration objectives presented in Table II.

| Calibration item | Optical Paths |
| --- | --- |
| RAD-1 | OP2, OP3, OP4 |
| RAD-2 | OP2 |
| GEO-1 | OP1 |
| GEO-2 | OP1, OP2, OP5 |
| SPE-1 | OP1, OP3, OP5 |
| SPE-2 | OP1 |
| SPE-3 | OP3, OP5 |

We summarize in Table IV the OP used to produce measurements associated with calibration items defined in Table II. A main collimating mirror ("collimator", Figure 3) was used to collimate test beams for OP1/2/5 and produce beams large enough to cover the full entrance of the instrument. This collimator was a heritage of the previous setups (see Section I C); its characteristics (focal length: 4 m and diameter: 400 mm[7]) put strong constraints on the OP1/2/5 calibration setup design. A rotating mirror was used to select OP among OP1/2/5. OP3 was located close to the TVC window in the new cubic unit, without the use of numerous optics. A retractable mirror was used to switch from OP1/2/5 to OP3. OP4 was located in the TVC and MAJIS was observing it using hexapod movements.

### B. TVC thermal equipment

As discussed previously (see section I C), one of the major challenges of the MAJIS calibration facility development was to manage the thermal environment, in particular inside the TVC. A new TVC thermal environment was therefore developed. It was intended to take in charge of three constraints: (1) reproduce low temperatures reached by MAJIS in space, (2) reduce infrared fluxes related to thermal contributions that may enter MAJIS aperture, and (3) create a light source (OP4) for IR radiometry.

MAJIS's main cooled elements are its OH and its IR detector. Passive cooling with a radiator exposed to space will be used around Jupiter[1]. This will result in a detector temperature of about 90 K and an OH temperature of about 130 K (see table III). In the TVC (Figure 5), IR detector thermal control was obtained through the use of a cryocooler, while the OH low temperature was obtained not through the dedicated radiator mounted but by attaching conductive copper braids cooled by liquid nitrogen to the OH structure. A water cooling system was also used to set MAJIS main electronics at ambient temperatures (20°C). The liquid nitrogen flux was used to cool a shield composed of aluminum plates surrounding MAJIS. This radiative shielding was used to facilitate MAJIS thermal control and reduce ambient infrared fluxes. This shield was set on the hexapod, thus it was mobile. In front of MAJIS, a 1.3 m long copper baffle (painted black inside) was used to connect the MAJIS aperture to the TVC window. The baffle cross section was decreasing from the TVC window to the inner, MAJIS facing aperture which size was 160 mm × 110 mm. The baffle geometric properties were chosen to ensure that the scan mirror surface was fully illuminated for all arrival light angles corresponding to the FOV (Table I). This baffle was surrounded by a copper front shield with a second aperture leading to OP4. The front shield, baffle, and OP4 were also cooled (see Figure 5). Flexible connections between fixed and mobile elements were obtained with Multi-Layer Insulation (MLI) (Figure 3); MLI was also placed all around the shields and baffle.

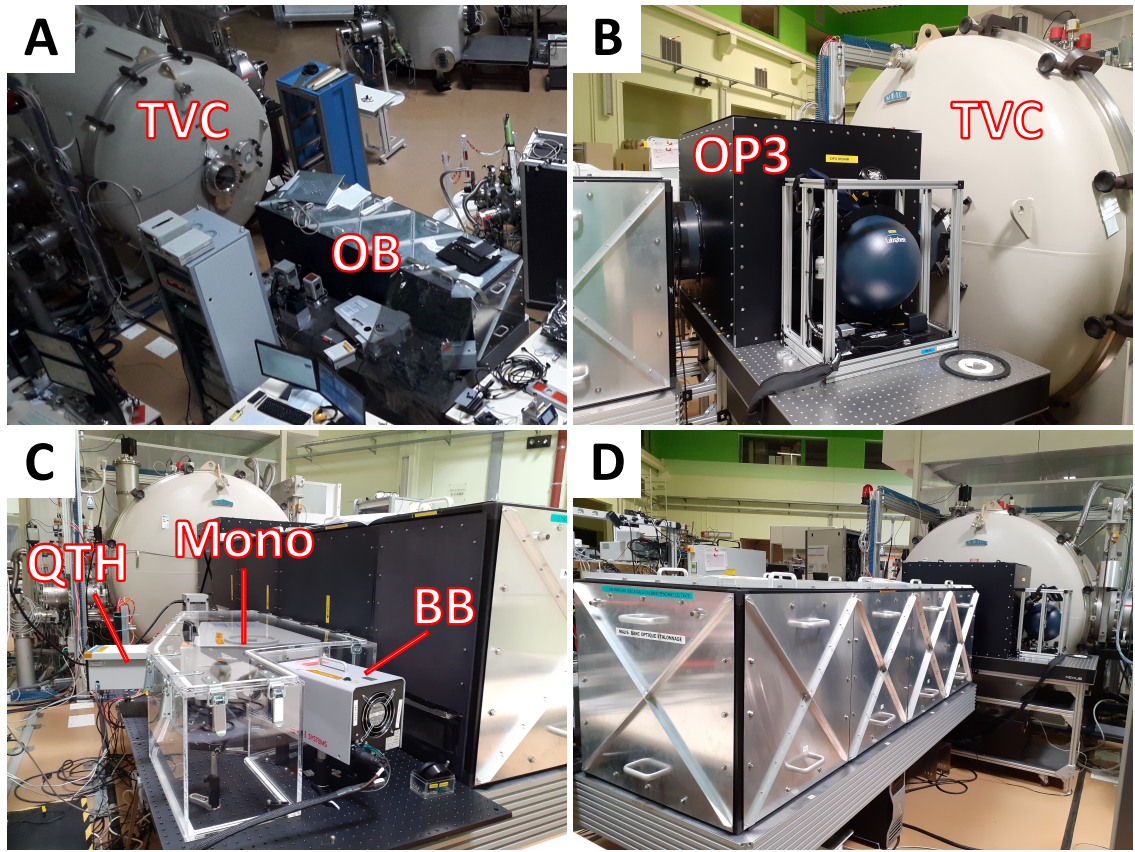

FIG. 4. External views of the Optical Bench (OB). (A) Top view of the TVC and OB in the former configuration of the setup as used for SIMBIO-SYS/BepiColombo in 2015, except that the OB is disconnected from the TVC window. (B) Zoom on the new cubic unit inserted between the TVC and the rest of the OB to accommodate OP3. (C) Front view of the OB, with the new transparent flushed cavity; OP1 elements are indicated: the monochromator ("Mono"), the QTH light source, and the black body light source ("BB"). (D) Back view of the OB, with the OP3 cubic unit and the TVC in perspective.

C. Hexapod compliance

The main mechanical requirements inside the TVC were related to the movements of the hexapod above which MAJIS OH was mounted. Details about this hexapod can be found in the publication dedicated to the SIMBIO-SYS/BepiColombo calibration campaign[7]. Prior calibration, we checked that hexapod performance was adapted to MAJIS calibration requirements. We summarize this analysis in the next paragraph and in Table V.

Hexapod movements were used (1) to point MAJIS optical axis toward either the direction of OP1/2/3/5 in the optical bench (OB) or OP4 in the TVC, (2) to maintain a given source within FOV while MAJIS scan mirror was employed, (3) to center light sources at different positions within MAJIS FOV, and (4) to position light source out of the FOV for straylight estimate. Table V contains the hexapod performance requirements in terms of absolute precision, relative precision, and stability. These requirements are compared to actual performance measured by the hexapod manufacturer for a mass of 145 kg representative of MAJIS and the mobile shield (see Section II B). Most movements during calibration consisted in rotations as rotation reproduces a variation of pointing of MAJIS. Absolute precision concerns the capability for the hexapod to reach a position from any starting point: rotation requirements were linked to the need for an accurate coupling between MAJIS boresight and source direction. Absolute precision near the edges of the hexapod operating area was lower, but needs were limited to low sampling foreseen for the external straylight exploration (typically, minimum steps in the 0.1 - 0.5° range). Relative precision was linked to the capability of the setup to reach a position from a nearby one. Relative precision needs were higher but can be satisfied through successive incremental steps. Translation needs were less demanding as translation movement correspond to the coupling between MAJIS entrance pupil and incoming collimated flux width from either OP1/2/3/5 (the same collimator was shared) or OP4 inside the TVC. Finally stability performance was compared to expected longest acquisition time. All characteristics of hexapod performance was evaluated to be compliant with needs for the calibration campaign.

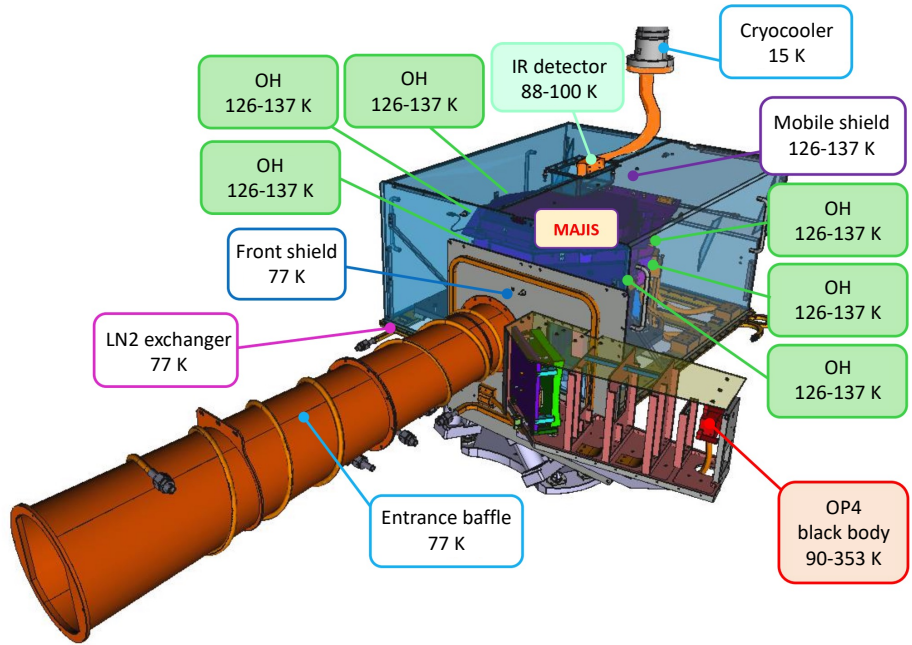

FIG. 5. Three-dimensional view of the thermal equipment surrounding MAJIS inside the TVC and operative temperatures. IR detector and OH thermal control were obtained through the use of a cryocooler and liquid nitrogen (LN2), respectively (explored values according to Table III). MAJIS was surrounded by a mobile shield that followed hexapod moves. A fixed entrance baffle (surrounded by a front shield) was used to link the MAJIS aperture and the TVC window. Temperature control was obtained using LN2 for the shield and baffle; mobile shield temperature was connected to that applied to the OH. Actual temperatures of part of these large elements may have been 10 K higher than set points during operations. The components of OP4 were also cooled using LN2, resulting in operatives temperatures for the OP4 black body (i.e., the light source of OP4) between 90 K and 353 K (80°C).

TABLE V. Hexapod performance compared to calibration requirements.

| Parameter | Requirement | Rationale | Calibration item | Performance |
|---|---|---|---|---|
| Rotation absolute precision (nominal) | < 75 µrad | IFOV/2 | GEO-1 | [15 - 60] µrad |
| Rotation absolute precision (edge) | < 2 mrad | Straylight steps | RAD-2 | [40 - 200] µrad |
| Rotation increment | 30 µrad | 1/5 IFOV | GEO-12 | Resolution < 3 µrad, repeatability < 5 µrad |
| Rotation stability | < 3 µrad over 4 s | 10% of 1/5 IFOV during highest $t_{int}$ | GEO-12 | < 3 µrad over 1 min |
| Translation absolute precision | < 750 µm | < 1% of entrance pupil | RAD-1 | < 40 µm |
| Translation increment | None | | RAD-1 | Resolution < 2 µm, repeatability < 5 µm |
| Translation stability | < 750 µm over 4 s | < 1% of entrance pupil, highest $t_{int}$ | RAD-1 | < 0.4 µm over 1 min |

### D. Optical path 1

OP1 was designed as a configurable source for the evaluation of the spectral and geometric performance (items GEO and SPE in Table II). Spatially, OP1 was producing a small spot, i.e., its spatial dimensions were adjustable but always lower than the FOV. Dimensions were typically between one and 20 IFOVs (see Table VI for precise dimensions). The spectral range of the source could vary from monochromatic adapted to SPE items, to polychromatic adapted to GEO items.

A schematic description of the components and optical paths for OP1 is available in Figure 6, and a picture of some components is shown in Figure 4C. OP1 was composed of the following elements, in that order:

- a source of light (selected among two, each coupled with its own optics);
- a monochromator;
- three plane mirrors (M3, G3 rotating mirror, and G2);
- a collimating mirror G1;

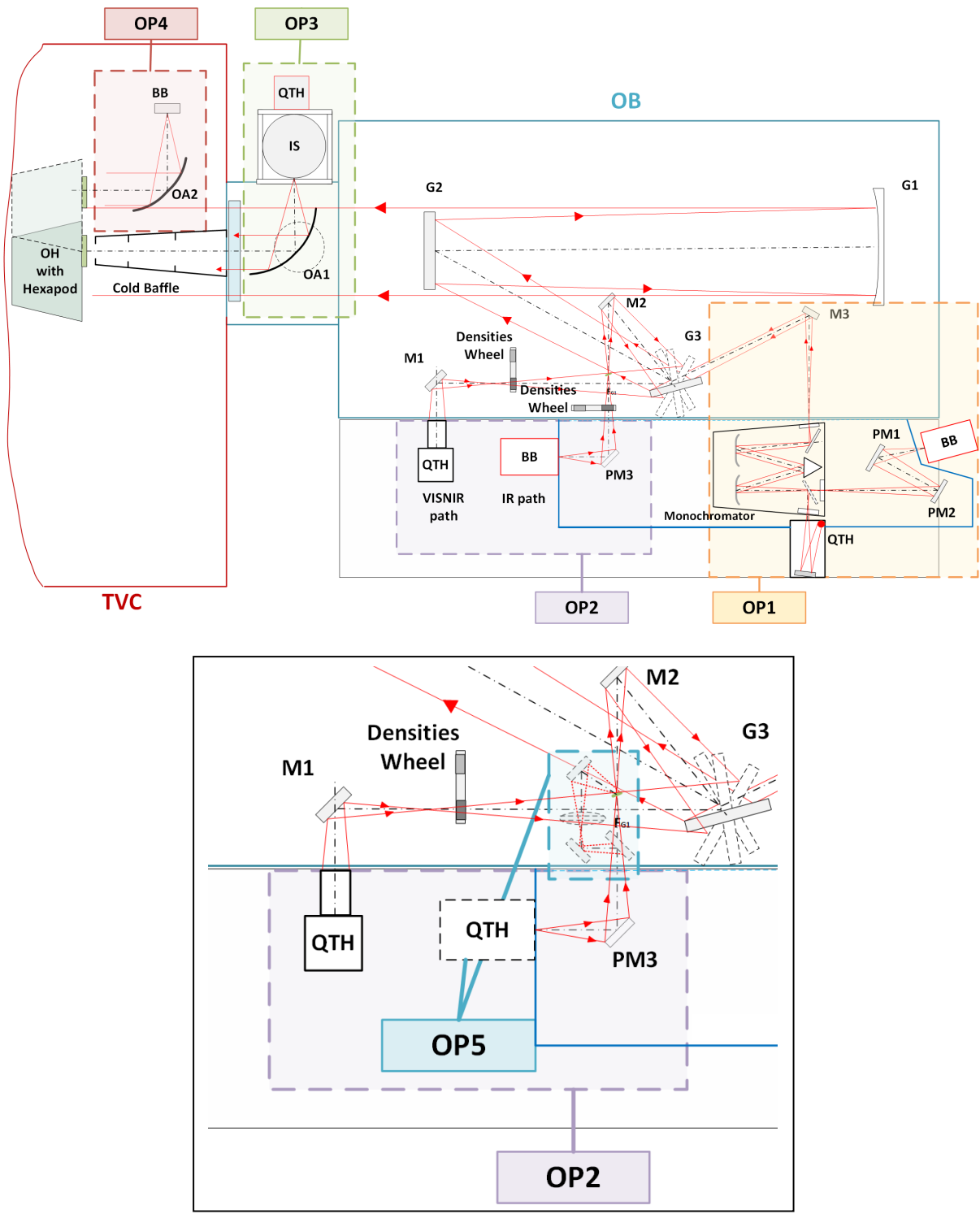

FIG. 6. Details of the five "Optical Paths" (OP) used for ground calibration at the IAS facility. OP are numbered from 1 to 5, and elements specific to a given OP are highlighted with colored rectangles with dashed contours. OP1, 2, and 5 share several optical elements, including the main collimating mirror G1, a plane mirror G2, and a rotating mirror G3 for OP selection. "M" is used to indicate a plane mirror, "PM" corresponds to off-axis parabolic mirrors and "OA" to off-axis conic mirrors. IS stands for Integrating Sphere and OH for Optical Head (of MAJIS). Light sources are indicated with "QTH" for Quartz Tungsten-Halogen lamps and "BB" for Black Bodies. OP2 and OP5 could not be operated simultaneously: the top scheme represents the configuration of the bench with OP2, while the bottom zoomed scheme shows the bench modified to implement OP5, with a movable unit installed in the optical bench (OB, dashed blue square) and the OP2 BB replaced by a QTH. OP3 included a retractable mirror OA1 that was mobilized to use either OP3 or OP1/2/5. The transparent nitrogen flushed box that contained some OP1 and OP2 elements is indicated with a solid blue line. OP4 was located in the TVC, and the OH of MAJIS was directed toward OP4 using hexapod movments.

- the TVC window.

The monochromator had two entries connected to the following two sources of light:

- A Quartz Tungsten Halogen lamp (QTH) lamp with an adjustable power, manufactured by *HORIBA*, covering the wavelength range of the VISNIR channel (Table I). Light is emitted through the heating of a tungsten filament at $\approx$ 3400 K for the maximum power level (100 W). Lamp irradiance peaked at $\approx$ 800 nm, with $\geq$ 15% of this maximum over [0.49 - 2.35] µm. Light emitted by the filament was concentrated with an off-axis parabolic mirror included in the QTH lamp delivered by the manufacturer. A diffuser (frosted glass) was added inside the QTH lamp, right after the filament (and before the parabolic mirror), to obtain a spatially homogeneous source. The QTH lamp output was placed in front of the first monochromator entrance (Figure 4C, Figure 6).

- A black body (manufactured by *CI Systems*, model SR200) with a variable temperature of up to $\approx$ 1200 K corresponding to a maximum of the Planck function at 2.4 µm, close to the lower wavelength edge of the IR channel at 2.27 µm. The black body specifications are: a short term stability (<1 h) of 0.25°C and a long term stability (>1 h) of $\pm$ 0.4°C, an absolute temperature accuracy of $\pm$ 2°C, and a resolution of 1°C. Injection in the second entrance of the monochromator (Figure 4C) was performed using two off axis parabolic mirrors (PM1 and PM2 on Figure 6). These mirrors (manufactured by *Edmund optics*) were coated with gold (emissivity < 2% over the IR range).

The monochromator was manufactured by *HORIBA* (model iHR 550). The expected maximum spectral resolution was 0.3 nm over the 0.5-2.3 µm range and 0.6 nm over the 2.3-5.6 µm range, and the expected absolute accuracy was $\pm$ 0.2 nm. These values must be compared to the accuracy requirement of the spectral calibration items ("SPE" in Table II): 0.7 nm and 1.3 nm for VISNIR and IR, respectively. The monochromator was equipped with three gratings, that could be selected using a motorized turret, so as to cover the whole MAJIS spectral range (see Table VII for details). A wheel with four broadband filters was present at the output of the monochromator to remove second orders. The monochromator could be used at zero in order to produce polychromatic light.

Spatially, the output of the monochromator was a slit with a length of 12 mm. The monochromator slit was aligned with the MAJIS slit. The relative magnification of OP1 to MAJIS is 0.06; the slit extent thus corresponded to a geometrical image size of 20 nominal pixels. The slit width was settable. For spectral items requiring a given spectral resolution (SPE items in Table II), the slit width was calculated for each wavelength according to manufacturer instructions and was comprised between 0.3 and 1.3 mm (see Table VII for details). It was possible to set gratings at zero order to obtain polychromatic outputs with wavelength ranges comparable to the initial light sources, which were adapted to evaluate geometric items ("GEO" in Table II) over large wavelength ranges at once. Larger values of slit width up to 7 mm were then possible. In order to obtain a disk source, an iris diaphragm with a variable diameter was added right after the output of the monochromator. This iris, manufactured by *SmarAct*, was used with diameters from 0.9 mm to 2 mm. The minimal value of 0.9 mm corresponded to 1.5 nominal pixels on MAJIS. The various combinations of slit widths and iris apertures used with OP1 are summarized in Table VI.

### E. Optical path 2

OP2 was primarily designed as a powerful source to estimate straylight coming from out-of-FOV directions. It was also used for other radiometric measurements, notably to compensate for a problem encountered with OP3 during the campaign (see Section IV), as well as for a few geometric measurements. A very high flux was required to detect slight straylight contributions originating several degrees away from the boresight (five order magnitude contrast, see Table II). This flux was attenuated while approaching the FOV using density filters: a filter with an optical density $D$ reduces the flux by $10^{-D}$. The value of $D$ can vary with wavelength around an average value $\overline{D}$.

A schematic description of the components and optical paths for OP2 is shown in Figure 6. Similarly to OP1, two different sources (QTH and black body) could be used for OP2, roughly corresponding to the VISNIR and IR channels respectively. The following elements succeeded each other, with some notable differences between the VISNIR and IR channels discussed thereafter:

- an initial source of light;

- a first mirror;

- density filters mounted on rotating wheels;

- two or three plane mirrors (M2 for IR only, G3 rotating mirror, and G2);

- a collimating mirror G1;

- the TVC window.

The OP2 section adapted to VISNIR was composed of a QTH lamp with an adjustable power of up to 240 W (manufactured by *Newport*) with a relative spectral behavior comparable to that of the QTH lamp employed in OP1. The QTH lamp (composed of a filament and an optic to concentrate light) was directly connected to the main structure of the optical bench (OB) toward a plane mirror (M1 on Figure 6). Then light went through two density filter wheels before reaching rotating mirror G3 to join the same path as for OP1. The first density filter wheel was equipped with an empty slot and a slot with a filter with an average optical density $\overline{D} = 2$. The second density wheel was equipped with an empty slot and a slot with a density filter $\overline{D} = 3.5$ obtained by combining a density filter $\overline{D} = 1$ and a density filter $\overline{D} = 2.5$. It was thus possible to reach an equivalent density filter $\overline{D} = 5.5$ by combining both wheels adequately.

The OP2 section adapted to IR was composed of the same model of the black body as used for OP1 (temperature up to $\approx$ 1200 K). This black body was connected to the transparent nitrogen flushed cavity (Figure 4C) where an off axis parabolic mirror was located (PM3 on Figure 6). Then light went through a density wheel inside the main cavity of the OB. The wheel contained three employed slots equipped with: no filter, a density filter $\overline{D} = 1$, and a density filter $\overline{D} = 3$, respectively. The light was then reflected on mirror M2 toward rotating mirror G3. From there the optical path was again similar to that of OP1.

Spatially, OP2 spot size and shape in the image plane was a disk with a diameter of $\approx$ 34 nominal pixels for the black body (IR), and a rectangle of about $26 \times 16$ nominal pixels for the QTH (VISNIR).

Finally, a half disk hole was also implemented in a slot of a rotating wheel to create an object with a well-defined edge useful for evaluating some GEO items (Table II).

TABLE VI. Size configurations of OP1. OP1 size was controlled by the width of the monochromator output slit and/or the circular iris aperture diameter that was set by indicating an angular position from 0 (close) to 100° (fully open). Slit length (along-slit) was 12 mm. Image sizes along and across correspond to lengths/widths/diameters calculated in the MAJIS focal plan with the 0.06 theoretical relative magnification between OP1 and MAJIS.

| Configuration | Slit width (*mm*) | Iris (°) | Iris diameter (*mm*) | Image along (*nominal pixels*) | Image across (*nominal pixels*) |
|---|---|---|---|---|---|
| Large slit | 7 | 100 | 24 | 20 | 11.7 |
| Thin slit | 1 | 100 | 24 | 20 | 1.7 |
| $\lambda$-slit | Table VII | 100 | 24 | 20 | 0.7 - 3.7 |
| Small spot 1 | 1 | 6 | 0.9 | 1.5 | 1.5 |
| Small spot 2 | 7 | 7 | 1.2 | 2 | 2 |
| Large spot 1 | 7 | 9 | 1.7 | 2.9 | 2.9 |
| Large spot 2 | 7 | 10 | 2.0 | 3.3 | 3.3 |

TABLE VII. Monochromator properties selected for spectral calibration. Three gratings were used to cover the MAJIS wavelength range. Slits size corresponds to the size of both the entrance slit and the exit slit of the monochromator, set to the same value. It was calculated to provide a fixed FWHM of the output spectral beam corresponding to $\Delta\lambda/5 = 0.7$ nm (VISNIR) and 1.3 nm (IR) respectively (see Table II). Grating unit is grooves per millimeter. Values for minimum and maximum wavelengths of each grating range are indicated.

| Grating ($gr.mm^{-1}$) (*Channel*) | Wavelength (*μm*) | Dispersion ($nm.mm^{-1}$) | Slits size (*mm*) |
|---|---|---|---|
| 1200 | $\lambda_{min}$ = 0.5 μm | 1.34 | 0.52 |
| (*VISNIR*) | $\lambda_{max}$ = 1.2 μm | 0.88 | 0.79 |
| 600 | $\lambda_{min}$ = 0.9 μm | 2.71 | 0.26 |
| (*VISNIR*) | $\lambda_{max}$ = 2.3 μm | 1.87 | 0.37 |
| 300 | $\lambda_{min}$ = 1.9 μm | 5.39 | 1.34 |
| (*IR*) | $\lambda_{max}$ = 5.5 μm | 2.70 | 0.48 |

### F. Optical path 3

OP3 was a full-FOV flat field source primarily adapted to the VIS-NIR channel radiometry. It was also used for some spectral calibration items. It was located in the optical bench (OB) unit directly connected to the TVC window (Figure 4). This made it possible to reduce the optical path length in the OB below the 2 m requirement so as to avoid possible spectral perturbations caused by the residual presence of water and carbon dioxide gas (see section I E). The optical design was conceived to limit as much as possible the number of optics so as to facilitate the assessment of the absolute radiance of the whole OP. It was composed of the following elements (Figure 6):

- an Integrating Sphere (hereafter IS) to which a 150 W QTH source was connected;
- a retractable off-axis conic aluminum mirror ("OA1") to collimate the output of the sphere;
- the TVC window.

The IS and its QTH source were manufactured by *Labsphere*. The QTH was used at a fixed electrical power, while variation of the light radiance of the IS was obtained using a variable shutter placed between the QTH and the IS. The sphere was calibrated by the manufacturer prior to delivery for five shutter levels over the VISNIR channel range. The absolute radiance uncertainties were evaluated to be < 1% between 0.5 and 2.3 μm and < 2% at 2.4 μm for shutter positions 0% (no attenuation), 21%, and 58%. For a shutter at 66%, the uncertainty was ≤ 1% between 0.5 and 2.0 μm and increased to 5% at 2.4 μm. For a shutter at 76%, the uncertainty was in the 2-5% range between 0.5 and 1.7 μm, 9% at 2 μm, and then increased strongly to 60% at 2.4 μm. An assessment based on the IS internal coating reflectance and QTH emitted flux showed that the OP3 setup would provide sufficient flux up to 2.8 μm so as to cover also 0.5 μm of the IR channel as required in the overall specifications of the setup discussed in section I C. The QTH lamp's stability was guaranteed for 50 h by the manufacturer, and its use was recorded to ensure that this threshold was not exceeded. In addition, two sensors were present inside the IS to monitor the flux during the calibration campaign.

The sphere diameter was 30.5 cm, and the output diameter was 10.2 cm. The uniformity of this output was evaluated by the manufacturer through a spatial mapping test during which the IS output was divided into 89 elements. Deviations from uniformity were measured to be ≤ 1% for all elements, a prerequisite for the entire OP3 to be compliant with the 1% relative requirement of Table II detailed in section I B.

During the campaign, the nitrogen flush was sometimes stopped or replaced by a carbon dioxide flush in the OB cubic unit dedicated to OP3. This procedure was put in place to acquire spectral measurements (SPE items) that take benefit from the presence of water and carbon dioxide well-defined spectral features (Figure 2).

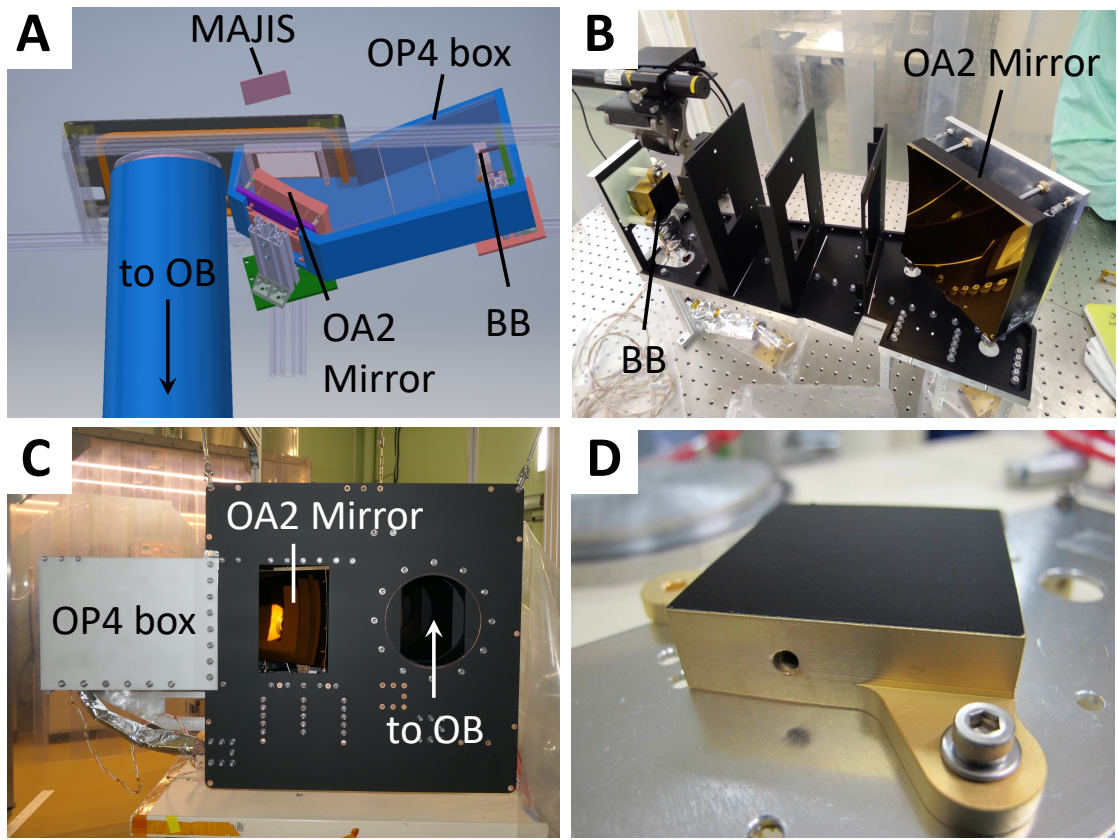

FIG. 7. Views and diagram of the radiometric IR light source "OP4". (A) Overall design of OP4 and its localization in the TVC. OP4 black body ("BB"), OA2 mirror, and enclosing box, are indicated. The position of the entrance of the MAJIS baffle looking at OP4 is represented ("MAJIS"). The main entrance baffle that links the TVC to the optical bench (OB) and thus to the other OPs is also shown. (B) View of the interior of the OP4 box while under test, showing the series of three baffles between the blackbody and the mirror. (C) View from MAJIS position, showing front shield opening toward the OB (OP1-3, 5), front shield opening toward OP4 mirror, and OP4 closed box. (D) Zoom on the OP4 blackbody, composed of a thick gold coated copper plate covered by a dark adhesive high emissivity element.

### G. Optical path 4

OP4 was a full-FOV flat field source primarily adapted to the NIR channel radiometry, but also to the VISNIR channel radiometry for wavelengths greater than ≈ 1.8 µm. It was entirely designed and manufactured at IAS for MAJIS. It was located in the TVC. This made it possible to reduce the thermal background contribution to negligible values up to a wavelength of 5.6 µm, and also to remove any potential issue related to residual atmospheric contributions, more prominent over the IR channel range compared to the VISNIR one (see Figure 2).

OP4 was composed of the following elements (Figure 7):

- a custom-made black body;
- a succession of three baffles;
- an off-axis conic mirror (OA2).

The custom-made black body was a square copper plate, 50 mm wide and 15 mm thick, with a gold coating. One face of the plate was covered with a dark adhesive aluminum foil provided by manufacturer *Acktar*, model *Metal Velvet*. This created a diffusive surface with a high emissivity/low reflectance over the visible and infrared wavelength range of MAJIS. The plate was connected to a thermal regulation system composed of (i) a connection to the liquid nitrogen network for cooling and (ii) a heater resistor to control temperature and obtain a radiometric IR light source. The accessible temperature range was 90-353 K.

The OA2 mirror, manufactured by *Pleiger*, was gold-coated, with a characterized reflectance spectrum between 98% and 99% over the 2-6 µm range. A set of three baffles was inserted between the mirror and the black body (Figure 7B), and all elements were located in an "OP4 box" bound to the front shield (Figure 7A). MAJIS could look toward OP4 through an opening in the front shield (Figure 7C). OP4 mirror, baffles and box walls were cooled similarly to the shield surrounding MAJIS in the TVC (see Figure 5).

### H. Optical path 5

OP5 was the optical path dedicated to the observation of solid samples and reference materials. The principle of the OP5 channel was to illuminate opaque samples and observe them in reflection. OP5 was composed of the following elements (Figures 6 and 8):

- a 100 W QTH source;
- a parabolic mirror (PM3);

TABLE VIII. Samples and reference surfaces used with OP5. The last column indicates their origin: collections of university laboratories *Institut de Planétologie et d'Astrophysique de Grenoble* (IPAG) and *Institut d'Astrophysique Spatiale* (IAS), or products from commercial suppliers *Space Resource Technologies* (SRT) and *Labsphere* (details about these latter references surfaces have been documented previously[12]; WCS stands for *Wavelength Calibration Standard*).

| Number | Nature | Type | Origin |
|---|---|---|---|
| 1-1 | Calcite | powder pellet | IAS |
| 1-2 | Dunite | powder pellet | IPAG |
| 1-3 | Kaolinite | powder pellet | IPAG |
| 1-4 | Sub-bituminous coal | powder pellet | IPAG |
| 1-5 | Spectralon 99% | calibration standard | Labsphere |
| 1-6 | WCS | calibration standard | Labsphere |
| 2-1 | BoPET (above mirror) | calibration standard | IAS |
| 2-2 | Biotite | rock fragment | IAS |
| 2-3 | Serpentine | rock fragment | IAS |
| 2-4 | Gypsum | powder pellet | IAS |
| 2-5 | CI meteorite analogue | powder pellet | SRT |

- a series of four optics located in the movable OP5 unit (two mirrors M4 and M5, a lens, and a mirror M6);
- a rotating wheel with six slots (two different wheels were used in turn);
- three plane mirrors (M2, G3 rotating mirror, and G2);
- a collimating mirror G1;
- the TVC window.

This channel was not permanently installed in the bench, as it shared parts of the optical path of OP2 (see Figure 6). Setting up OP5 required (i) opening the bench, to place a movable unit blocking the OP2 VISNIR optical path, and (ii) replacing OP2's black body source with a QTH source.

Three types of material were placed in the rotating wheel slots: (i) powder pellets, (ii) rock fragments, and (iii) calibration standards. Elements placed in the rotating wheels are listed in Table VIII. Details about the composition and spectral properties of the samples are provided in a companion paper[23]. Illumination was limited to a QTH lamp, covering the VISNIR channel and the beginning of the IR channel, due to the high thermal emission of samples at ambient temperature (see Figure 1).

## III. SETUP PERFORMANCE CHARACTERIZATION

### A. Test devices

Setup actual performance was characterized mostly before the calibration campaign between 2019 and 2021. A few additional measurements were obtained after the campaign. No ground instrument with performance comparable to that of MAJIS was available. This characterization thus relied on the use of several test devices, each adapted to a given specification type and/or OP (e.g., spectral accuracy of OP1 and OP3, spatial homogeneity of OP3 and OP4, etc.) Performance characterization was not exhaustive and was sometimes limited to just a few wavelengths or flux levels. Used devices are summarized in Table IX.

Spatial performance in the focal plane was evaluated using "cameras" (i.e., a 2D pixel array), with 15 or 16 µm pixel size, placed at the focal plane of a 250 mm focal length lens. The resulting configuration was close to that of MAJIS (240 mm focal length and 18 µm unbinned pixel size, see Table I), enabling a spatial sampling similar to that of MAJIS. Two cameras were used:

- a *ANDOR* DU897-BV camera: 512 x 512 pixels, 16 µm pixel size, 0.4-1 µm range, hereafter "ANDOR";
- a *INFRATEC* ImageIR8300 camera (without its default lens): 640 x 512 pixels, 15 µm pixel size, 2-5.7 µm range, hereafter "INFRATECH".

The spatial performance of the collimated beam that illuminated the MAJIS entrance baffle (i.e., the beam "at pupil level") was evaluated with single-pixel detectors placed on movable devices with two used configurations:

- a *Thorlabs* PDA10JT-EC detector (2.5 to 5.5 µm wavelength range) placed on a translation table located outside the TVC, between collimating mirror G1 and the TVC window (see Figure 6);
- a *Thorlabs* PDA20H detector (1.5 to 4.5 µm) or a *OPHIR* PD300-IR detector (0.7 to 1.8 µm) placed on the hexapod inside the TVC.

Spatial performance in the focal plane was also evaluated with a mock-up of MAJIS (Figure 9) placed under vacuum in the TVC, above the hexapod, and inside the thermal equipment detailed in Figure 5. This mock-up was designed to mimic MAJIS and its primary objective was to test the TVC thermal environment. It was equipped with a MCT detector from manufacturer *Infrared Associates* that was 2 mm wide, cooled to 77 K, with a wavelength range from 2 µm to 6 µm (detector sensitivity thus covered the IR range and a small part of the VISNIR range).

Spectral performance was evaluated with a *Spectral Evolution* PSR+ 3500 spectrometer (hereafter "PSR"). This device covered the 0.35 to 2.5 µm range with a spectral resolution (FWHM) between 2.8 and 8 nm depending on wavelength. Its wavelength reproducibility was 0.1 nm. Flux was collected with an optical fiber at various locations of the setup (see Table IX).

Typical characterization steps and key measurements required for subsequent data processing are presented for each OP (except OP5 for which a dedicated paper exist[23]) in the following sections.

### B. OP1

#### 1. Spatial performance

First, we evaluated the shape, size and homogeneity of OP1 sources in the focal plane for various OP1 source size configurations (see Table VI). Figure 10 shows the spatial extent of the source when the OP1 configuration corresponds to the maximum source size (slit fully open). These measurements provided an estimate of the actual source size and made it possible to quantify the spatial variability of the source's flux. Measurements are consistent with a 20 x 12 nominal pixels size in the MAJIS focal plane, as expected from theory (see Table VI). Along- and across-slit flux variations were estimated. In the representative example of Figure 10 obtained in the VISNIR configuration of OP1, we observe that the flux decreases along and across slit from a maximal value at the center. Across-slit, the variation is limited to ±1% over 4 MAJIS nominal pixels: flux could thus be considered constant for the smaller slit size used for spectral characterizations (see Tables VII and VI). Along-slit, the variability is more pronounced, with a progressive decrease from 100% at the center to 50% at $\pm 10$ nominal pixels from the center. A constant flux was however not required along-slit as the along-slit dimension of OP1 was exploited to characterize several pixels at once, but each with its own relative across-slit or spectral variations. Figure 11 illustrates the OP1 characterization when source size is limited by the iris. Flux profiles are illustrated for two configurations. We show in the left panel the along and across slit profiles when the iris was opened to its minimum value (0.9 mm diameter corresponding to "small spot 1" in Table VI, a configuration frequently used for VIS-NIR measurements): these profiles show that the small image spot is symmetrical with a FWHM of $1.4 \pm 0.1$ nominal pixels consistent with the expected geometric size of 1.5 nominal pixels (Table VI). In the right panel, we illustrate another configuration corresponding to a larger iris aperture adapted for IR ("large spot 1" in Table VI). Two wavelengths of the IR channel are shown: the profile is not significantly modified between both with a similar FWHM of $3.4 \pm 0.1$ nominal pixels. This value is comparable, although slightly larger, to the expected geometrical size of 2.9 nominal pixels indicated in Table VI.

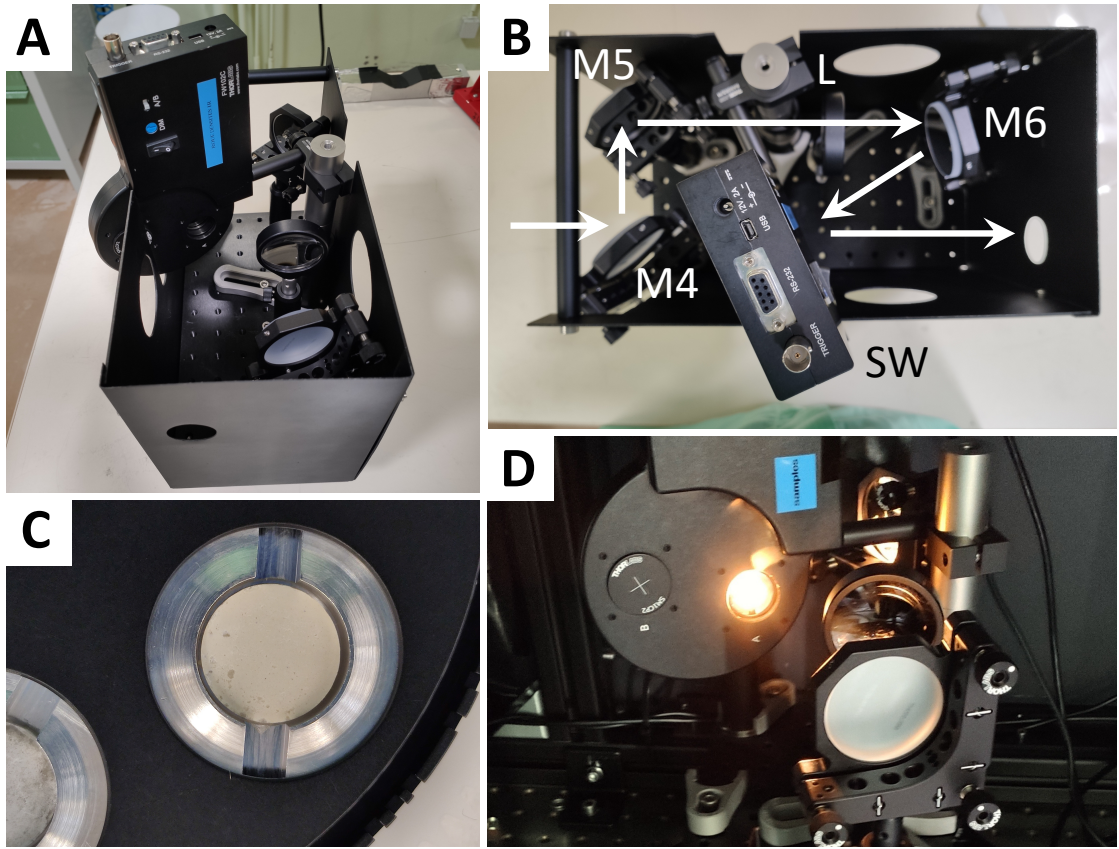

FIG. 8. Views of the samples optical path "OP5". (A) Overview of the OP5 unit, placed in the optical bench (OB). (B) Top view of the interior of the OP5 unit with the three mirrors (M4, M5 and M6), the lens (L), and the samples wheel (SW). The direction of light within the unit is indicated with arrows. (C) Zoom on a sample holder within the wheel, with the kaolinite sample (see Table VIII) mounted on it. Part of the dunite sample is also apparent on the bottom left corner. (D) Preliminary test inside the OB of the optical elements of the OP4 unit, with a QTH source illuminating a sample.

Second, the spatial extent and homogeneity of the collimated beam that illuminated the MAJIS entrance baffle were evaluated with the the following configuration of OP1: The source was the black body at 1200 K, the input and output slit widths were 5 mm and 1 mm, respectively, the grating was 300 $gr.mm^{-1}$, and 7 wavelengths between 2.5 and 5.5 µm were tested. The PDA10JT-EC detector (Table IX) mounted on a translation table was used for this test, with monitoring of the flux every millimeter. Observed flux spatial fluctuations over the whole beam were characterized by a standard deviation of about 1%. The test was also used to adapt baffles located in the optical bench (OB) to fully illuminate the entrance of the TVC baffle.

### 2. Spectral performance

OP1 was the sole OP capable of sending monochromatic light. Its spectral performance was evaluated with the PSR spectrometer. We compare in Figure 12 the OP1 prescribed wavelength and the actual central wavelength observed with the PSR spectrometer. Observed variations are $\leq \pm 2$ nm and can be attributed to the reproducibility of the monochromator in operative conditions, which notably included automatic changes of gratings. The PSR spectral resolution was significantly larger than that of OP1, by a factor of 4-10 (PSR FWHM was between 2.8 and 8 nm over the VISNIR range). It was thus not possible to precisely check the theoretical FWHM of OP1 presented in Table VII. Nevertheless, measurements were conducted and showed that the FWHM of OP1 observed by the PSR was comparable to that of the PSR only, which was compatible with the expected low theoretical FWHM of OP1.

TABLE IX. Summary of test devices used to characterized the actual performance of the calibration setup. Length indicated in the "Main optics" column are focal lengths. Please refers to Figure 6 for localisation. Tested OPs with each devices are indicated in the last column.

| Name | Detector size | Wavelengths | Main optics | Localisation | OP |
|---|---|---|---|---|---|
| ANDOR | 512 × 512, 16 μm pixels | 0.4 - 1 μm | 250 mm lens | Between G1 mirror and TVC window, or inside TVC above hexapod | OP1,2,3 |
| INFRATECH | 640 × 512, 15 μm pixels | 2.0 - 5.7 μm | 250 mm lens | Between G1 mirror and TVC window, or inside TVC above hexapod | OP1,2,3 |
| PDA10JT-EC | Single pixel | 2.5 - 5.5 μm | - | Between G1 mirror and TVC window on a translation table | OP1 |
| PDA20H | Single pixel | 1.5 - 4.5 μm | - | Inside TVC above hexapod | OP1 |
| PD300-IR | Single pixel | 0.7 - 1.8 μm | - | Inside TVC above hexapod | OP1 |
| PSR | Single pixel | 0.3 - 2.5 μm | Fiber optic | Between OA1 mirror and TVC window, or at monochromator / density filters outputs | OP1,3 |
| Mock-up | Single pixel (2 mm) | 2 - 6 μm | 250 mm mirror | inside TVC above hexapod | OP3,4 |

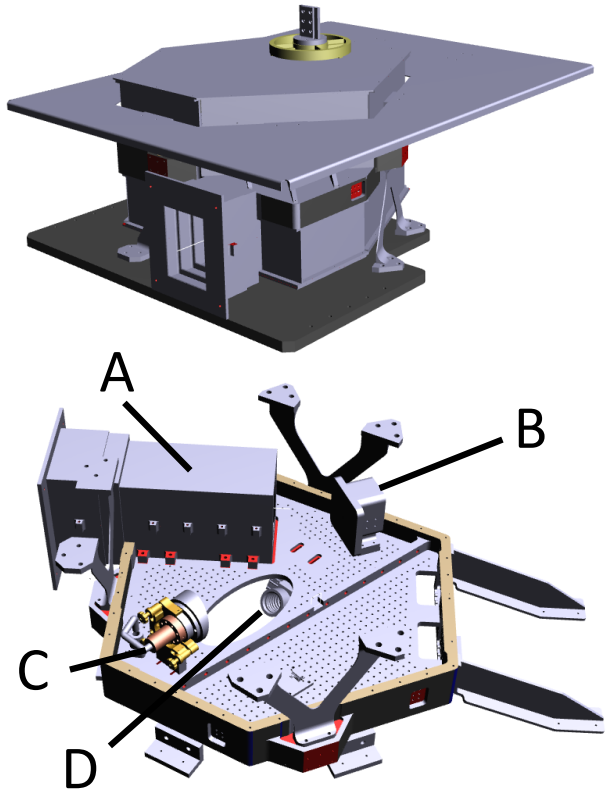

FIG. 9. MAJIS mock-up used to evaluate the optical performance of OP3 and OP4. Top: global view with hood. Bottom: detailed view of the interior with (A) entrance baffle; (B) square plan mirror 100 mm wide; (C) off-axis parabolic mirror with a focal of 250 mm; (D) focal plane unit, composed of a MCT detector, a cryogenic chopper, and a baffle.

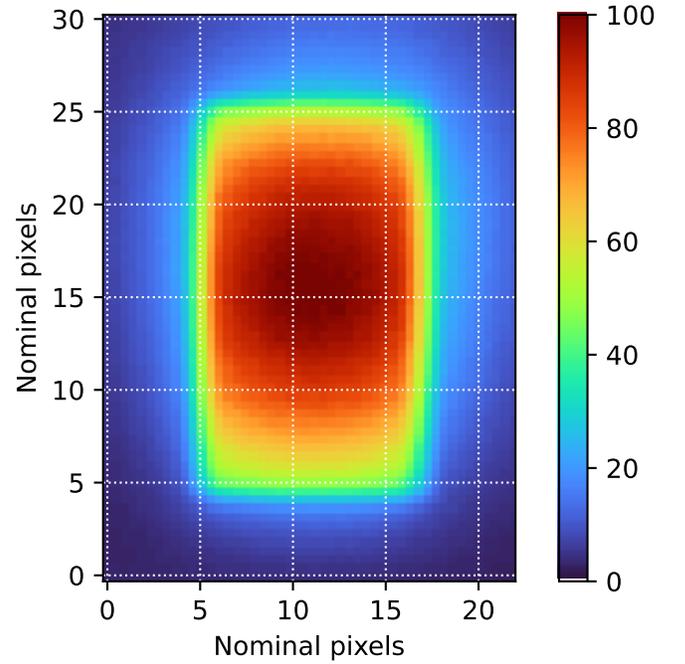

FIG. 10. Spatial extent and flux variability of OP1 in the focal plane for the "large slit" configuration (no source size reduction, see Table VI) and a typical VISNIR wavelength (grating is 600 $gr.mm^{-1}$ and $\lambda = 1.2$ μm; QTH source is used). Coordinates are converted to MAJIS nominal pixels. Intensity normalized to 100 is shown.

In addition, we performed spectral scans, with a 1 nm sampling, of a portion of the atmospheric water spectrum located between 2620 and 2720 nm. The output slit of the monochromator was directly imaged with the INFRATECH camera. We did not detect any spectral variability along-slit. Considering the shape of the spectral features over that range, spectral shifts > 1 nm would have been detected. We also compared the shape of the feature with the actual water transmission using HITRAN[21]: features were in agreement and we estimate that an absolute shift > 2 nm would have been identified. This is consistent with the measurements of Figure 12.

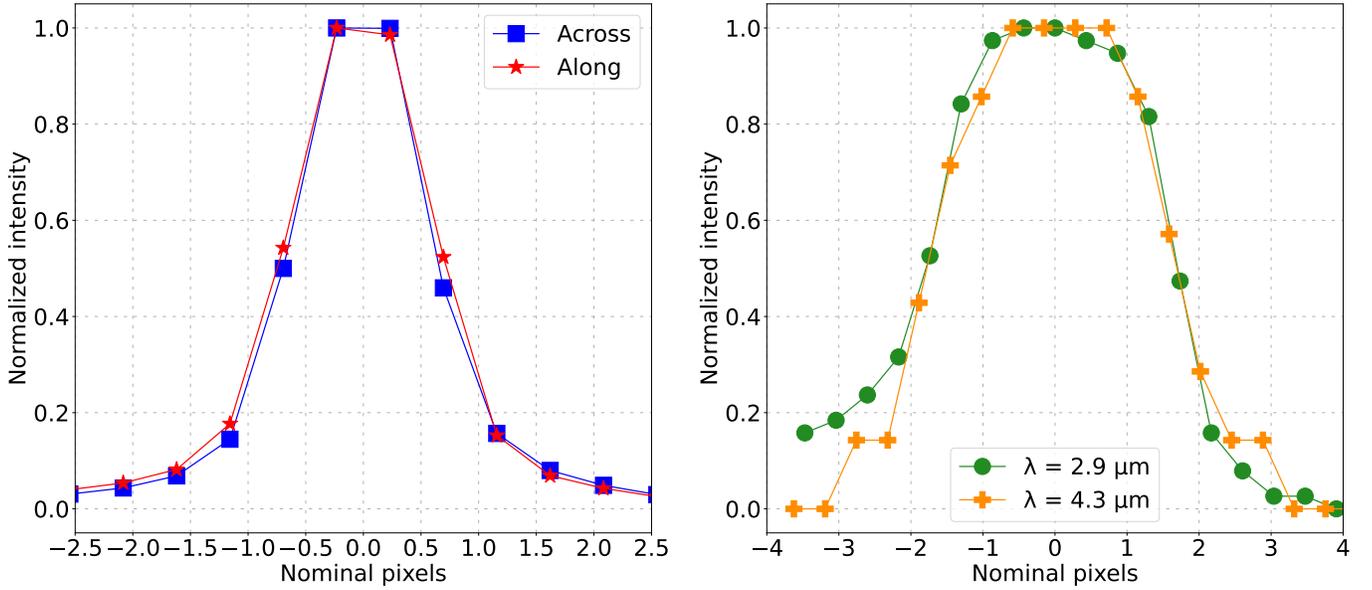

FIG. 11. Spatial intensity profiles of OP1 in the focal plane when iris is used (disk shape). Left panel: "Small spot 1" configuration as defined in Table VI; across-slit (blue squares) and along-slit (red stars) profiles are shown for $\lambda = 1.1$ μm and 600 $gr.mm^{-1}$ grating. Right panel: "Large spot 1" configuration; along-slit profiles for wavelengths $\lambda = 2.9$ μm (green dots) and $\lambda = 4.3$ μm (yellow crosses) are shown (grating is 300 $gr.mm^{-1}$).

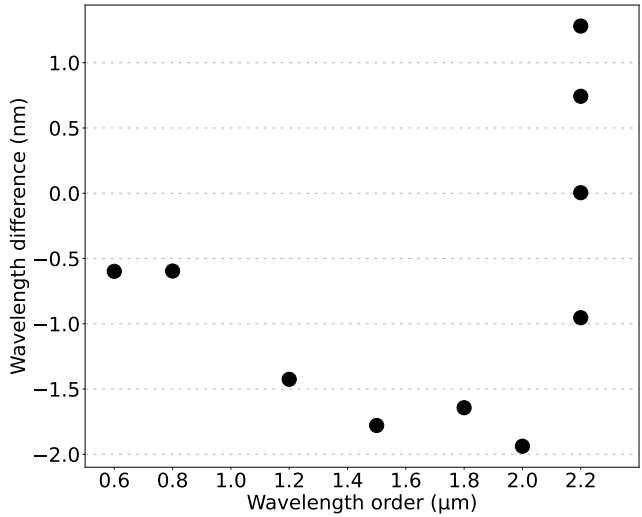

FIG. 12. Evaluation of OP1 absolute wavelength precision and reproducibility. The wavelength difference between the measured value and the ordered value is shown over the VISNIR range. Observed variations are $\leq \pm 2$ nm. The 4 measurements obtained at the same wavelength (2240 nm) directly illustrate the reproducibility in operative conditions, that was degraded compared to expected manufacturer performance (see Section II D). All observed variations with wavelength can actually be attributed to reproducibility rather than constant systematic biases. Two different monochromator gratings were used to obtain these measurements (see Table VII).

### C. OP2

One of the specificities of OP2 was that it relied on the use of density filters, so as to comply with the $10^{-5}$ flux contrast requirement for straylight evaluation (see Table II). The properties of density filters used to attenuate the powerful OP2 flux were checked as a function of wavelength with the PSR spectrometer. Radiance spectra with density filters ($I(\lambda)$), and without ($I_0(\lambda)$), were collected, so as to calculate optical density $D(\lambda) = -log(I(\lambda)/I_0(\lambda))$. Results for the three density filters of the OP2 section adapted to the VISNIR channel are shown in Figure 13. The optical density of each filter actually significantly varies with wavelength around the average expected value $\overline{D}$: $0.7 \leq D(\lambda)/\overline{D} \leq 1.5$. This characterization will be necessary to interpret MAJIS straylight measurements as a function of wavelength.

Similar measurements as for OP1 were conducted to check OP2 spatial properties. Data obtained with MAJIS during the calibration campaign with the VISNIR configuration notably revealed the presence of a faint parasitic secondary source located at a distance lower than MAJIS FOV from the primary source. This secondary source was caused by some optical bench light leakage around the density filter wheels and was characterized in greater detail right after the calibration campaign, in November 2021, with the ANDOR camera. Results are illustrated on Figure 14 where we can observe that the rounded shape of the edge of the density filter wheel delimits the secondary source. The primary source corresponds here to the lamp going through two density filters (total optical density $\overline{D} = 3.5$), while the secondary source does not go through any filter. The secondary source flux level is lower than 3.5% of the maximum flux level of the primary source with $\overline{D} = 3.5$: the parasitic image intensity is actually only $10^{-5}$ that of the initial source, which explains why it was not initially detected. It also implies that this contribution will be negligible while analyzing MAJIS data obtained without density filters, i.e. most data for which the source is out-FOV. This parasitic source is non negligible essentially in reference measurements obtained with the OP2 source in-FOV.

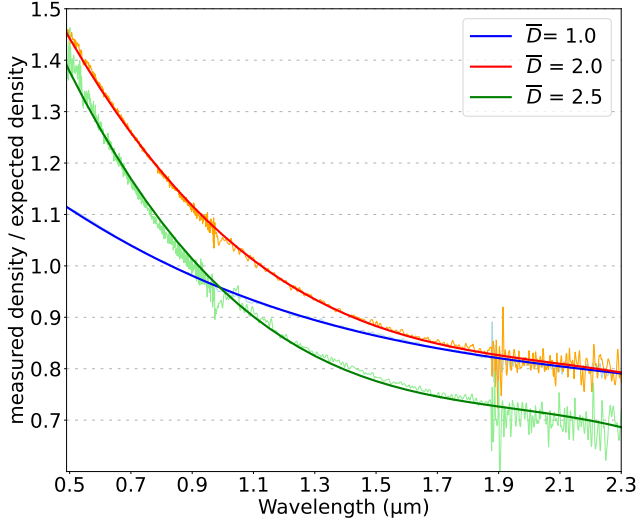

FIG. 13. Spectral variability of the optical density for the three filters used in the OP2 section adapted to the VISNIR channel. The ratio of the measured density $D(\lambda)$ over the expected average density $\overline{D}$ is shown.

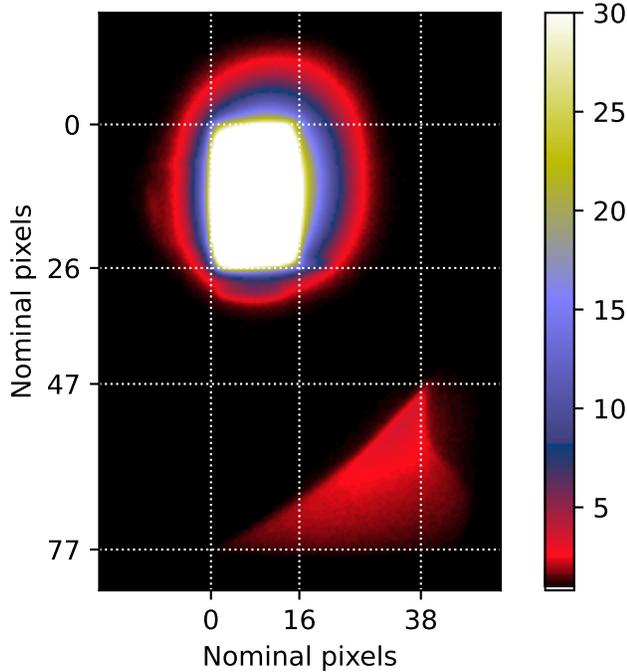

FIG. 14. Spatial properties of OP2 in the focal plane for the VIS-NIR configuration. Measurements revealed the presence of a parasitic secondary source (bottom right) due to light leakage on the side of the filter wheel. Intensity normalized to 100 at the center of the primary source (top left) is shown (the color bar stops at 30, i.e., values greater than 30 are in white). The primary OP2 source included a total optical density $\overline{D} = 3.5$, while light from the parasitic secondary source was not attenuated by density filters.

## D.  OP3

OP3 was designed to be a radiometrically calibrated flat field covering the full-FOV of MAJIS over the VISNIR range, plus the early IR range. First, the extent and homogeneity of the collimated beam at the level of the MAJIS baffle entrance was evaluated with the single-pixel detector PDA20H and various wavelength filters placed inside the TVC on the hexapod. Translations with a 2 mm spatial sampling were conducted. Most variations of this flux were observed to be within ± 5%.

Then, we evaluated the spatial homogeneity of OP3 in the focal plane, for which strong requirements have been specified (Table II) as OP3 is a flat field source. The characterization was notably conducted with the mock-up (Figure 9), hence this homogeneity evaluation included the effects of all optical elements of OP3. The mapping of the OP3 source was obtained by rotating the viewing direction of the mock-up with the hexapod (Figure 15): these measurements showed a spatial variability of the flat field consistent with the $\leq 1\%$ requirement over the whole MAJIS FOV (i.e., over 3.4° along-slit, see Table I) at low spatial sampling (approximately one point every 40/25 nominal pixels in the across/along slit direction respectively). Additional measurements were also conducted with the ANDOR camera and showed that the variability of OP3 at MAJIS nominal sampling was also $\leq 1\%$.

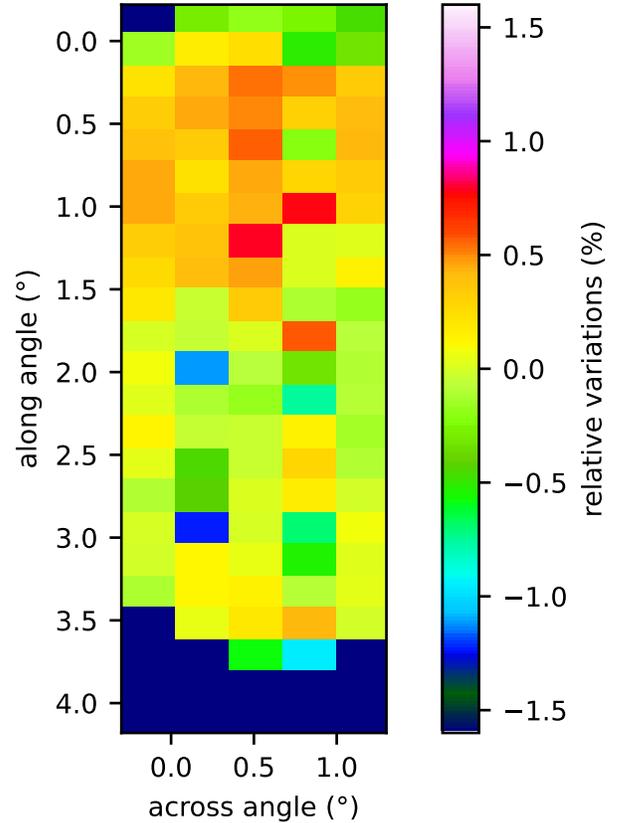

FIG. 15. Homogeneity of OP3 flat field. The central section of OP3 was mapped in the focal plane with the mock-up of MAJIS (Figure 9) placed above the hexapod inside the TVC. Relative variations with respect to the median are shown.

Finally, the radiometric properties were analyzed in detail as OP3

was initially designed to be used for absolute radiometry (Figure 16). However, due to an issue with the positioning of the OP3 retractable mirror ("OA1") during the calibration campaign, the actual flux absolute level was different from the expected flux. Nevertheless, OP3 was still usable for relative measurements (flat field, linearity, and spectral variations of the ITF), and other multifunctional OPs (OP2 and OP4) were used to scale OP3 flux to its absolute value[24].

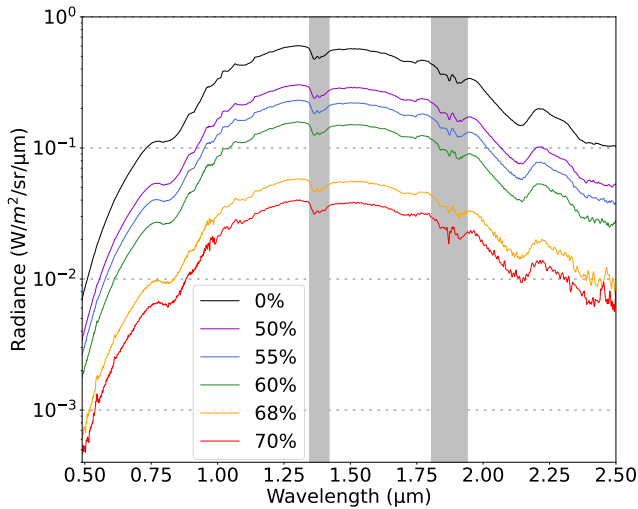

FIG. 16. Evaluation of the spectral radiance of OP3 for various IS shutter values (0% corresponds to a fully open shutter). Radiance was measured with the PSR spectrometer placed between TVC window and OA1 mirror (see Figure 6) and modulated by the previously measured TVC transmission (about 90%[7]). The actual QTH lamp used during the calibration campaign was employed, with the same current of 3.5 A. Test measurements were conducted without nitrogen flush, hence atmospheric water vapor lines (gray area) can be observed. Note that the actual absolute flux level during the campaign was different, due to a mechanical issue with the positioning of the OP3 mirror (see discussion in Section IV).

### E. OP4

OP4 was a flat field source used for absolute radiometry, similar to OP3, but located inside the TVC and covering the IR channel wavelength range plus part of the VISNIR range. It was specially designed for this calibration campaign (see Section II G). The light source was a black body surface composed of a copper plate covered with a dark adhesive (see Figure 7D). The radiometric precision of OP4 relied on the knowledge of the actual temperature and emissivity of this assembly (absolute value at a given point and spatial homogeneity). Temperature was controlled by a high-precision probe ($\Delta T/T \leq 1\%$) connected to one point of the copper plate. Spatial homogeneity tests were conducted with the mock-up (Figure 9), i.e., under vacuum and over the foreseen operative temperature range. The homogeneity of OP4 in the focal plane (Figure 17) was compliant with requirements with observed variations $\leq 1\%$. Finally, the spectral properties of the black coating were measured, with an observed reflectance between 1.1% and 1.7% over the 1-5 μm range at 300 K, in agreement with its expected high emissivity.

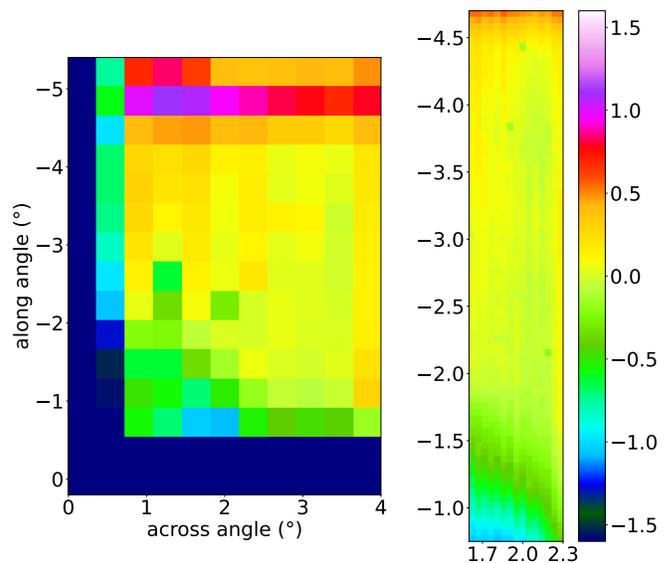

FIG. 17. Homogeneity of OP4 flat field. OP4 was mapped in the focal plane with the mock-up of MAJIS placed above the hexapod inside the TVC. OP4 black body temperature was 50°C. Left: global assessment of the almost entire source at low spatial sampling. Right: high spatial sampling (approximately one point every 0.05°, i.e. 6 IFOV) characterization of a subset of the source. Relative variations in % with respect to the median (similarly as for Figure 15) are shown with the same color code for both panels.

## IV. OVERVIEW OF MEASUREMENTS OBTAINED WITH MAJIS

Measurements were obtained with this setup and MAJIS from August 30th to September 17th, 2021, over 15 working days. This time interval was significantly shorter than initially foreseen due to planning constraints on the delivery date of MAJIS. This led us to adapt and optimize the measurements. In addition, a new component of the VISNIR MAJIS straylight was identified for the first time during the IAS calibration campaign (Figure 18); this had a significant impact on the course of the measurements, as part of the time was devoted to characterizing it. The main measurements carried out in this context are summarized in Table X. Three temperature configurations were tested in agreement with Table III: a nominal case with $T_{OH}$ = 126 K and $T_{IR\ detector}$ = 88 K, a hot case with $T_{OH}$ = 137 K and $T_{IR\ detector}$ = 96 K, and two additional values of $T_{IR\ detector}$ (92 K and 100 K) at $T_{OH}$ = 137 K.

During OP3 and OP4 measurements (Table X, Nos. 25-38), the whole FOV was illuminated at once with a homogeneous source. Measurements were repeated for about 10 integration times and 10 source flux levels (Table X) so as to sample precisely MAJIS radiometric response. These key measurements were acquired to produce the dataset needed to derive the instrument transfer function in all conditions, which includes the assessment of linearity, saturation and flat field response (item "RAD-1" of Table II). Figure 19 illustrates the actual background signal observed by MAJIS with these two OP when sources were off, which is well in line with expectations presented in Figure 1. For OP3, we can see that the thermal background at ambient temperature is indeed enough to saturate the IR channel beyond 4 μm (i.e., over about 40% of the IR wavelength range) at low integration time (85 ms). For OP4, we can see that the specific thermal environment created in the TVC (Figure 5) was ap-propriate to suppress any measurable thermal contributions. OP3 and OP4 source sizes in the focal plane were significantly larger than the MAJIS slit. This was used to perform additional measurements over various parts of these sources by rotating the hexapod (Table X, Nos. 27, 34, and 37) so as to mitigate hypothetical flat field inhomogeneity not identified during the setup characterization steps (Figures 15 and 17). In fact, OP4 observations by MAJIS turned out to contain a decrease in intensity by up to 17%[24], essentially localized on one edge of the FOV. This decrease could be due to a vignetting issue related to an inappropriate relative positioning of MAJIS and OP4. Hexapod rotations were also used to point MAJIS boresight toward the center of OP3 and OP4 for various MAJIS scan mirror angular positions (Table X, Nos. 26, 33, and 36). OP3 encountered an issue related to some mechanical instabilities in the positioning of the retractable OA1 mirror (Figure 6), which precluded reliably using OP3 alone to assess the absolute radiance in the VISNIR. An alternative was found with the use of OP2 with its black body source (Table X, Nos. 17-19). This was made possible by the fact that the temperature reachable by OP2 black body was much higher than that of OP4 (1200 K vs 350 K), as OP2 main goal was to have a powerful source for straylight, while OP4, adapted for IR radiometry, was expected to cover only the longest wavelengths of the VISNIR channel for cross comparison between OP4 and OP3.

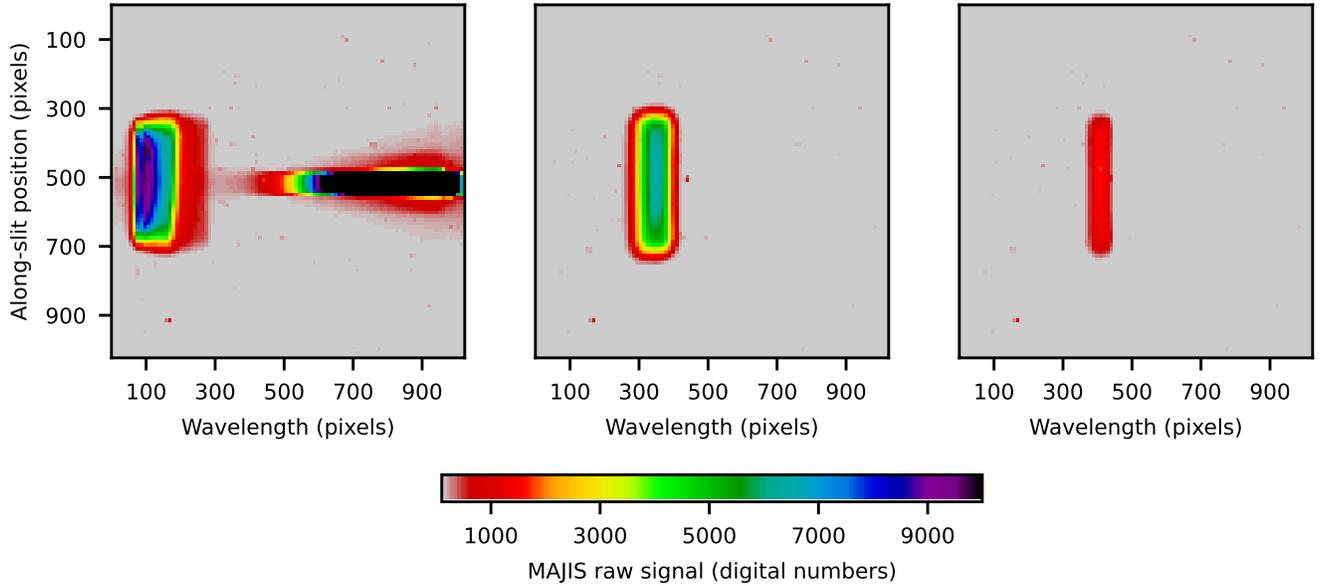

FIG. 18. MAJIS VISNIR observations of OP2, illustrating the straylight detected during the on-ground calibration campaign. (Left) OP2 is in-FOV, at the center of the slit (y ≈ 500), and the increasing flux of the 250°C OP2 black body is perceptible for x ≥ 400; the straylight corresponds to the vertical rectangle between x ≈ 50 and x ≈ 250 ($\lambda$ in the 0.6 to 0.9-1 µm range[3]). (Middle) OP2 is 0.6° away from the FOV and is no longer directly observable; the straylight remains, but it is narrower, shifted to longer wavelengths (now centered on x ≈ 350, i.e. $\lambda$ ≈ 1.1-1.2 µm) and attenuated in intensity. (Right) OP2 is 1° away from the FOV: the straylight is further shifted and attenuated. Considering all incoming angles present for an extended source then results in a significant straylight impact over the 0.6-1.3 µm range[24]. Coordinates are provided in unbinned pixels. Saturation is at ≈ 22500 DN. Measurements are extracted from sequence No. 21 of Table X.

OP2 was also used to explore straylight, its original main purpose. In the standard exploration sequence (Table X, No. 20), measurements were conducted with the following angular sampling beyond the slit (either along or across): 0.3°, 0.7°, 1.2°, 2°, 4°, 6.5°, and 9°. This exploration was done across-slit from three positions within FOV (-1.7°, 0° and 1.7°, i.e., bottom, center, and top of the slit) and along-slit from both the bottom and the top of the slit. In-FOV observations were obtained with the highest available optical density values ($\overline{D}$ = 5.5 for VISNIR and $\overline{D}$ = 3 for IR; see Section II E). Angles close to the slit were measured with both no density filters and intermediate optical densities ($\overline{D}$ = 2 for VISNIR and $\overline{D}$ = 1 for IR). OP2 was also used in an adapted way focused on the characterization of the VISNIR straylight identified during the campaign (Figure 18). The fact that this straylight was not identified in previous validation steps of the optical head of MAJIS by the Leonardo company is currently interpreted by a difference in the illumination conditions of MAJIS during tests conducted there: while the whole scan mirror was illuminated at IAS (Section II B), it was only partly illuminated during optical head validation at the Leonardo company (illumination was restricted to the entrance pupil). The IAS configuration is closer to the real illumination conditions of MAJIS that will be encountered at Jupiter, with the MAJIS entrance scan mirror fully illuminated. This straylight has some major impact on the instrument transfer function calculation[24] and interpretation of samples observed during the ground campaign with OP5[23]. Its precise behavior is complex and not yet fully understood. Its position in wavelength on the detector is a function of the angular position of the source in comparison to the field of view (Figure 18). This straylight contribution thus varies depending on source size (small sources like OP1 or OP2, vs large sources that extend beyond the FOV such as OP3 and OP4).

OP1 measurements were conducted to assess the spatial and spectral properties of MAJIS. OP1 measurements frequently consist of scans with steps of 1/5 of the resolution in spatial or spectral dimension to estimate the spatial point spread function (both across and along slit) as well as the spectral response function (items "GEO-2" and "SPE-2" of Table II). These measurements also enable boresight and FOV determination (item "GEO-1"), as well as spectral calibration and spectral smile evaluation (item "SPE-1"). Across-slit scans with a small source lead to unexpected observations with some signal measured significantly farther from the boresight compared to the theoretical slit FWHM of about 1.5 IFOV. Consequently, associated measurements were extended to ± 12-16 IFOV (Table X, Nos. 2, 4, 6, and 8). These observations, which are not in line with other measurements of the across-slit PSF[15], are not yet understood.

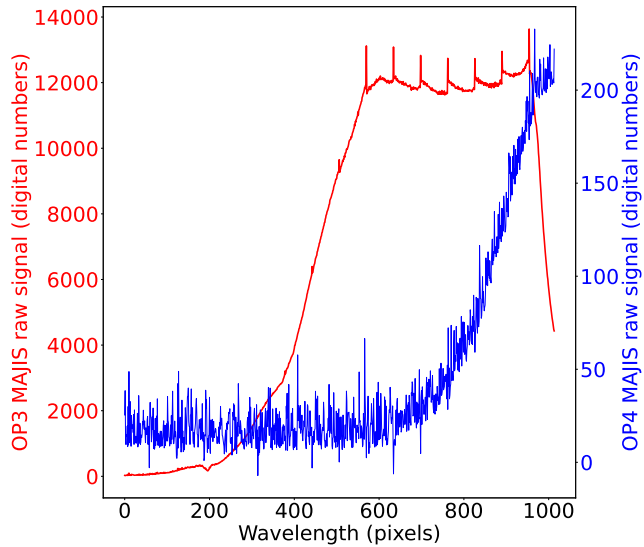

FIG. 19. Observed setup background flux with MAJIS IR channel during on-ground campaign for OP3 (red) and OP4 (blue). The integration time is 85 ms. OP3 measurement is extracted from sequence No. 25 of Table X with source off. OP4 measurement is extracted from sequence No. 32 while black body was regulated to 90K. Background corresponds to thermal flux at either ambient (red) or MAJIS OH (blue) temperature, in agreement with pre-campaign simulations (Figure 1). Coordinates correspond to unbinned pixels.

MAJIS VISNIR straylight illustrated for OP2 in Figure 18 was also strongly apparent in OP1 measurements. This complicated absolute spectral calibration measurements with the monochromator in the 0.6-1.3 µm range. Conducting such measurements was also a thorny issue in the IR channel due to the presence of the thermal background at ambient temperature (Figure 19). In addition, one must notice that OP1 absolute precision was slightly lower than expected (Figure 12). All this lead to an increase in the number of measurements performed with other OPs relying on reference materials, namely, OP5 with its solid samples and OP3 and OP2 for which dedicated measurements with water and $CO_2$ gas lines were conducted (Table X, Nos. 23, 24, 28, 30, and 31). To do so, the nitrogen flush was stopped to introduce ambient air with water vapor or $CO_2$ gas in the bench. Spectral lines identified in these measurements have proven to be highly reliable in performing the spectral calibration over the entire FOV of MAJIS[3]. Finally, one should notice that a slight spectral mismatch has been identified between these OP3 gas sample measurements and OP5 solid sample measurements[23], which may be related to a modification of the straylight contribution linked with the difference in illumination condition between OP3 (large source) and OP5 (smaller source).

## V. CONCLUSION

In this article, we presented and characterized the on-ground calibration facility developed and implemented for the MAJIS instrument aboard the JUICE mission. This setup was designed on the basis of an installation previously used to calibrate other space instruments. A number of technical innovations were conceived to adapt to MAJIS, whose spectral coverage extends up to 5.6 µm. One of the special features of the new setup was the use of light sources both inside and outside the thermal vacuum chamber. An absolute radiometric black body source was indeed designed for the occasion, and placed inside the chamber to eliminate parasitic thermal flux. A hexapod was used to point MAJIS toward either internal or external sources without chamber opening, so as to be able to carry out all geometric, spectral and radiometric characterization measurements within 15 days. At the design stage, it was decided to build a system that would enable key MAJIS features to be evaluated redundantly with several sources. This proved a wise choice to compensate for anomalies. The calibration campaign also saw the discovery of the presence of some MAJIS straylight in the VISNIR channel. This altered the course of several calibration steps and necessitated the prompt setting up of dedicated characterization sequences, which was feasible thanks to the versatile nature of the calibration setup. Overall, it was possible to acquire the dataset needed to calculate the radiometric instrument transfer function, perform the spectral and spatial calibration, and characterize the performance of MAJIS, as detailed in a series of companion papers[3,15,23,24].

## ACKNOWLEDGMENTS

The French contribution to MAJIS has been technically supported and funded by CNES – CONTRACT CNES – CNRS No. 180 117. The Italian contribution to MAJIS has been coordinated and funded by Italian Space Agency – CONTRACT No. 2021-18-I.0 and supported by the ASI-INAF Agreement No. 2023-6-HH.0.

The authors would like to acknowledge the numerous persons that have provided some help to design, build, characterize, manage, or operate the calibration setup, in particular Alessandra Barbis, Ferenc Borondics, Rosario Brunetto, Vincent Carlier, Sandrine Couturier, Bruno Crane, Karin Dassas, Marc Dexet, Jean-Pierre Dubois, Serge François, Leandro Gasparini, Brigitte Gondet, Ludovic Gonnod, Chloë Guallar, Cyrille Hannou, Véronique Hervier, Hong Van Hoang, Christian Ketchazo, Fabien Labanard, Paul Lami, François Langlet, Hugo Le Bars, Jean-Christophe Le Clech, Sandrine Maloreau, Magali Mebsout, Gilles Morinaud, Christine Nicolas, Léna Parc, Gilles Poulleau, Philippe Pradel, Claudia Ruiz de Galarreta, Christophe Sandt, Catherine Tamiatto, Leonardo Tommasi, Federico Tosi, Stephane Tosti, and Stéphane Trochet.

The authors also thank the two anonymous reviewers for their attentive proofreading.

## DATA AVAILABILITY STATEMENT

The data that support the findings of this study are available from the corresponding author upon reasonable request.

TABLE X. List of main measurements obtained with the on-ground setup and MAJIS. "Position" refers to the relative angular position of OP center compared to FOV center (along/across angles respectively). Various integration times ($t_{int}$) were tested [a]. "MAJIS scan" refers to MAJIS scan mirror pointing direction across FOV (see Table III). OP1 source sizes are defined in Tables VII and VI. "BB" refers to black body (and $T_{BB}$ is the associated temparature). The "Straylight list" of angular positions is detailed in the text. Most sequences include measurements of the setup background (source off) and MAJIS dark (closed MAJIS shutter). Corresponding calibration items from Table II are indicated.

| # | OP | TOH/TIR | Position (°) | $t_{int}$ (ms) | Other parameters | Items |
|---|----|---------|--------------|----------------|------------------|-------|
| 1 | 1 | 126 / 88 K | 0, ± 1.5 / 0 | 400 | Along-slit scan: ± 20 IFOV/5 steps (OP1 QTH, small spot 1) | GEO-12 |
| 2 | 1 | 126 / 88 K | 0, ± 1.5 / 0 | 400 | Across-slit scan: ± 80 IFOV/5 steps (OP1 QTH, small spot 1) | GEO-12 |
| 3 | 1 | 126 / 88 K | 0, ± 1.5 / 0 | 60 | Along-slit scan: ± 20 IFOV/5 steps (OP1 BB, small/large spot 2) | GEO-12 |
| 4 | 1 | 126 / 88 K | 0, ± 1.5 / 0 | 20 | Across-slit scan: ± 60 IFOV/5 steps (OP1 BB, small spot 2) | GEO-12 |
| 5 | 1 | 137 / 96 K | 0 / 0 | 400 | Along-slit scan: ± 20 IFOV/5 steps (OP1 QTH, small spot 1) | GEO-12 |
| 6 | 1 | 137 / 96 K | 0 / 0 | 400 | Across-slit scan: ± 80 IFOV/5 steps (OP1 QTH, small spot 1) | GEO-12 |
| 7 | 1 | 137 / 96 K | 0 / 0 | 60 | Along-slit scan: ± 20 IFOV/5 steps (OP1 BB, small spot 2) | GEO-12 |
| 8 | 1 | 137 / 96 K | 0 / 0 | 20 | Across-slit scan: ± 60 IFOV/5 steps (OP1 BB, small spot 2) | GEO-12 |
| 9 | 1 | 126 / 88 K | 0, ± 1.5 / 0 | 2000 | Spectral scan: ± 25 $\Delta\lambda$/5 steps ($\lambda$ = 0.9 μm, 1.4 μm, 1.9 μm) | SPE-12 |
| 10 | 1 | 137 / 100 K | 0, ± 1.5 / 0 | 2000 | Spectral scan: ± 25 $\Delta\lambda$/5 steps ($\lambda$ = 1.4 μm) | SPE-12 |
| 11 | 1 | 126 / 88 K | 0 / 0 | 2000 | Spectral scan: ± 25 $\Delta\lambda$/5 steps ($\lambda$ = 0.55 μm) | SPE-12 |
| 12 | 1 | 137 / 100 K | 0 / 0 | 2000 | Spectral scan: ± 25 $\Delta\lambda$/5 steps ($\lambda$ = 0.55 μm) | SPE-12 |
| 13 | 1 | 137 / 100 K | 0 / 0 | 2000 | Spectral scan: ± 25 $\Delta\lambda$/5 steps ($\lambda$ = 2.2 μm) | SPE-12 |
| 14 | 1 | 126 / 88 K | 0, ±1,5 / 0 | 2000 | Spectral scan: ± 25 $\Delta\lambda$/5 steps ($\lambda$ = 2.6 μm) | SPE-12 |
| 15 | 1 | 137 / 96 K | 0 / 0 | 2000 | Spectral scan: ± 25 $\Delta\lambda$/5 steps ($\lambda$ = 2.6 μm) | SPE-12 |
| 16 | 1 | 137 / 100 K | 0 / 0 | 1000 | Spectral scan: ± 25 $\Delta\lambda$/5 steps ($\lambda$ = 2.6 μm) | SPE-12 |
| 17 | 2 | 126 / 88 K | 0 / 0 | Full list | OP2 $T_{BB}$: 80, 100, 120, 250, 650 °C | RAD-1 |
| 18 | 2 | 126 / 88 K | 0, ± 1.5 / 0 | 100, 1000 | OP2 $T_{BB}$: 250, 350, 425, 500 °C | RAD-1 |
| 19 | 2 | 137 / 96 K | 0, ± 1.5 / 0 | Full list | OP2 $T_{BB}$: 120, 250, 350, 500, 600 °C | RAD-1 |
| 20 | 2 | 126 / 88 K | Straylight list | 85 | Straylight exploration up to ± 9° with various density filters | RAD-2 |
| 21 | 2 | 137 / 96 K | 0 / 0 to ± 1.6 (step every 0.2) | 1000 | Across-slit scan to characterize VISNIR straylight (MAJIS shutter open, $T_{BB}$: 250 °C; and MAJIS shutter closed, $T_{BB}$: 350 °C) | RAD-2 |
| 22 | 2 | 126 / 88 K | 0, ± 0.9, ± 1.5 / 0 | 12, 20 | MAJIS across-slit scan of the half-disk hole | GEO-2 |
| 23 | 2 | 126 / 88 K | 0 / 0 | 12 | No N2 flush (water vapor in), $CO_2$ added | SPE-3 |
| 24 | 2 | 137 / 96 K | 0, + 1.7 / 0 | 12 | No N2 flush (water vapor in), $CO_2$ added | SPE-3 |
| 25 | 3 | 126 / 88 K | 0 / 0 | Full list | OP3 shutter: 0, 50, 55, 60, 65, 68, 70, 72, 75, 77, 80% | RAD-1 |
| 26 | 3 | 126 / 88 K | 0 / 0 | 85, 200, 800 | MAJIS scan: -4°, -2°, 0°, 2°, 4° | RAD-1 |
| 27 | 3 | 126 / 88 K | 0 / ± 0.2, 0.4 | 85, 200, 800 | Across-slit scan of OP3 source | RAD-1 |
| 28 | 3 | 126 / 88 K | 0 / 0 | 20, 40, 80, 100 | No N2 flush (water vapor in); OP3 shutter 70% | SPE-3 |
| 29 | 3 | 137 / 96 K | 0 / 0 | Full list | OP3 shutter: 70% | RAD-1 |
| 30 | 3 | 137 / 96 K | 0 / 0 | 20, 40, 80, 100 | No N2 flush (water vapor in); OP3 shutter 70% | SPE-3 |
| 31 | 3 | 137 / 96 K | 0, ± 1.5 / 0 | 20, 40, 80, 100 | No N2 flush (water vapor in), $CO_2$ added; OP3 shutter 70% | SPE-3 |
| 32 | 4 | 126 / 88 K | 0 / 0 | Full list | OP4 $T_{BB}$: -100, -80, -50, -30, -10, 10, 30, 50, 80 °C | RAD-1 |
| 33 | 4 | 126 / 88 K | 0 / 0 | 100, 800 | MAJIS scan: -4°, -2°, 0°, 2°, 4° ($T_{BB}$: -50, -30, 30 °C) | RAD-1 |
| 34 | 4 | 126 / 88 K | 0 / ± 0.2, 0.4 | 100, 800 | Across-slit scan of OP4 source ($T_{BB}$: -50, -10, 30 °C) | RAD-1 |
| 35 | 4 | 137 / 96 K | 0 / 0 | Full list | OP4 $T_{BB}$: -50, -10, 30, 80 °C | RAD-1 |
| 36 | 4 | 137 / 96 K | 0 / 0 | 100, 800, 4000 | MAJIS scan: -4°, -2°, 0°, 2°, 4° ($T_{BB}$: -50 °C) | RAD-1 |
| 37 | 4 | 137 / 96 K | 0 / ± 0.2, 0.4 | 100, 800 | Across-slit scan of OP4 source ($T_{BB}$: -50 °C) | RAD-1 |
| 38 | 4 | 137 / 92, 100 K | 0 / 0 | 100, 400, 800, 1000, 2000 | OP4 $T_{BB}$: -30, 10 °C | RAD-1 |
| 39 | 5 | 126 / 88 K | 0, ± 1.5 / 0 | Full list | Samples and reference surfaces listed in Table VIII | SPE-3 |
| 40 | 5 | 126 / 88 K | 0 / 0 | 12, 40 | MAJIS imaging with scan mirror of "Calcite", "WCS" and "Serpentine" (Table VIII) | SPE-3, GEO-2 |

[a] The "Full list" of integration times was typically $t_{int}$ = 85, 120, 200, 400, 800 ms (1 MHz detector mode[24]) and $t_{int}$ = 800, 1000, 1500, 2500, 4000 ms (1 kHz detector mode[24]), with some variations (e.g., 100 and 2000 ms selected instead of 1500/2500 ms and 85/120 ms respectively; 4000 ms value not always measured; etc.)